\documentclass[10pt,journal,compsoc]{IEEEtran}
\ifCLASSOPTIONcompsoc
  \usepackage[nocompress]{cite}
\else
  \usepackage{cite}
\fi
\ifCLASSINFOpdf
  \usepackage[pdftex]{graphicx}
\else
\fi
\usepackage{amsmath}
\usepackage{physics} % for $\bra{...}$ instead of $\langle ... |$, $\ket{...}$ instead of $| ... \rangle$, \braket{...}{...}

\usepackage[colorlinks=true,linkcolor=blue,citecolor=blue,urlcolor=blue]{hyperref}

\usepackage{ifthen}

\newboolean{VersionForJournal}
\setboolean{VersionForJournal}{true}
\newboolean{IncludeAppendix}
\setboolean{IncludeAppendix}{false}
\newboolean{UseHighResolutionImages}
\setboolean{UseHighResolutionImages}{false}

\newcommand{\svdh}{state vector difference highlighting}
\newcommand{\vuni}{visual universality}

\usepackage{soul} % for \st{...} strikethrough text
\usepackage{color}
\newcommand{\removedtext}[1]{} % final version
\begin{document}

% Titles are generally capitalized except for words such as a, an, and, as,
% at, but, by, for, in, nor, of, on, or, the, to and up, which are usually
% not capitalized unless they are the first or last word of the title.
% Linebreaks \\ can be used within to get better formatting as desired.
% Do not put math or special symbols in the title.
\title{Visualizing Quantum Circuits: State Vector Difference Highlighting and the Half-Matrix}
%
%
% author names and IEEE memberships
% note positions of commas and nonbreaking spaces ( ~ ) LaTeX will not break
% a structure at a ~ so this keeps an author's name from being broken across
% two lines.
% use \thanks{} to gain access to the first footnote area
% a separate \thanks must be used for each paragraph as LaTeX2e's \thanks
% was not built to handle multiple paragraphs
%
%
%\IEEEcompsocitemizethanks is a special \thanks that produces the bulleted
% lists the Computer Society journals use for "first footnote" author
% affiliations. Use \IEEEcompsocthanksitem which works much like \item
% for each affiliation group. When not in compsoc mode,
% \IEEEcompsocitemizethanks becomes like \thanks and
% \IEEEcompsocthanksitem becomes a line break with idention. This
% facilitates dual compilation, although admittedly the differences in the
% desired content of \author between the different types of papers makes a
% one-size-fits-all approach a daunting prospect. For instance, compsoc 
% journal papers have the author affiliations above the "Manuscript
% received ..."  text while in non-compsoc journals this is reversed. Sigh.

\author{Michael~J.~McGuffin
        and Jean-Marc Robert% <-this % stops a space
\IEEEcompsocitemizethanks{\IEEEcompsocthanksitem M. McGuffin and J.-M. Robert are with \'{E}cole de technologie sup\'{e}rieure,
Montreal,
%Canada.
% Canada.\protect\\
% % note need leading \protect in front of \\ to get a newline within \thanks as
% % \\ is fragile and will error, could use \hfil\break instead.
% E-mail: michael.mcguffin@etsmtl.ca
%\IEEEcompsocthanksitem J.-M. Robert is with \'{E}cole
%de technologie sup\'{e}rieure, Montreal, Canada.
%\IEEEcompsocthanksitem O. Landon-Cardinal is with \'{E}cole
%de technologie sup\'{e}rieure,
%Montreal,
Canada.}% <-this % stops an unwanted space
\ifthenelse{\boolean{VersionForJournal}}{
\thanks{}
}{} % ifthenelse
}

% note the % following the last \IEEEmembership and also \thanks - 
% these prevent an unwanted space from occurring between the last author name
% and the end of the author line. i.e., if you had this:
% 
% \author{....lastname \thanks{...} \thanks{...} }
%                     ^------------^------------^----Do not want these spaces!
%
% a space would be appended to the last name and could cause every name on that
% line to be shifted left slightly. This is one of those "LaTeX things". For
% instance, "\textbf{A} \textbf{B}" will typeset as "A B" not "AB". To get
% "AB" then you have to do: "\textbf{A}\textbf{B}"
% \thanks is no different in this regard, so shield the last } of each \thanks
% that ends a line with a % and do not let a space in before the next \thanks.
% Spaces after \IEEEmembership other than the last one are OK (and needed) as
% you are supposed to have spaces between the names. For what it is worth,
% this is a minor point as most people would not even notice if the said evil
% space somehow managed to creep in.

\ifthenelse{\boolean{VersionForJournal}}{

% The paper headers
\markboth{}%
{McGuffin \MakeLowercase{\textit{et al.}}: Visualizing ...}
% The only time the second header will appear is for the odd numbered pages
% after the title page when using the twoside option.
% 
% *** Note that you probably will NOT want to include the author's ***
% *** name in the headers of peer review papers.                   ***
% You can use \ifCLASSOPTIONpeerreview for conditional compilation here if
% you desire.

}{} % ifthenelse

% The publisher's ID mark at the bottom of the page is less important with
% Computer Society journal papers as those publications place the marks
% outside of the main text columns and, therefore, unlike regular IEEE
% journals, the available text space is not reduced by their presence.
% If you want to put a publisher's ID mark on the page you can do it like
% this:
%\IEEEpubid{0000--0000/00\$00.00~\copyright~2015 IEEE}
% or like this to get the Computer Society new two part style.
%\IEEEpubid{\makebox[\columnwidth]{\hfill 0000--0000/00/\$00.00~\copyright~2015 IEEE}%
%\hspace{\columnsep}\makebox[\columnwidth]{Published by the IEEE Computer Society\hfill}}
% Remember, if you use this you must call \IEEEpubidadjcol in the second
% column for its text to clear the IEEEpubid mark (Computer Society jorunal
% papers don't need this extra clearance.)

% use for special paper notices
%\IEEEspecialpapernotice{(Invited Paper)}

% for Computer Society papers, we must declare the abstract and index terms
% PRIOR to the title within the \IEEEtitleabstractindextext IEEEtran
% command as these need to go into the title area created by \maketitle.
% As a general rule, do not put math, special symbols or citations
% in the abstract or keywords.
\IEEEtitleabstractindextext{%
\begin{abstract}
Existing graphical user interfaces for circuit simulators
often show small visual summaries of the reduced state of each qubit,
showing the probability, phase, purity, and/or Bloch sphere coordinates associated with each qubit.
These necessarily provide an incomplete picture of the quantum state of the qubits,
and can sometimes be confusing for students or newcomers to quantum computing.
We contribute two novel visual approaches to provide more complete information about small circuits.
First, to complement information about each qubit, we show the complete state vector,
and illustrate the way that amplitudes change from layer-to-layer under the effect of different gates,
by using a small set of colors, arrows, and symbols.
We call this ``state vector difference highlighting'', % ``state vector diffing''
and show how it elucidates the effect of Hadamard, X, Y, Z, S, T, Phase, and SWAP gates,
where each gate may have an arbitrary combination of control and anticontrol qubits.
Second, we display pairwise information about qubits (such as concurrence and correlation)
in a triangular ``half-matrix'' visualization.
Our open source software implementation, called MuqcsCraft,
is available as a live online demonstration
that runs in a web browser without installing any additional software,
allowing a user to define a circuit through drag-and-drop actions, and then simulate and visualize it.
\end{abstract}

% Note that keywords are not normally used for peerreview papers.
%
% Network visualization, tangible user interfaces, preregistered study
\begin{IEEEkeywords}
Quantum circuit, quantum algorithm, graphical user interface, GUI, state vector simulation, entanglement.
\end{IEEEkeywords}}

% make the title area
\maketitle

% To allow for easy dual compilation without having to reenter the
% abstract/keywords data, the \IEEEtitleabstractindextext text will
% not be used in maketitle, but will appear (i.e., to be "transported")
% here as \IEEEdisplaynontitleabstractindextext when the compsoc 
% or transmag modes are not selected <OR> if conference mode is selected 
% - because all conference papers position the abstract like regular
% papers do.
\IEEEdisplaynontitleabstractindextext
% \IEEEdisplaynontitleabstractindextext has no effect when using
% compsoc or transmag under a non-conference mode.

% For peer review papers, you can put extra information on the cover
% page as needed:
% \ifCLASSOPTIONpeerreview
% \begin{center} \bfseries EDICS Category: 3-BBND \end{center}
% \fi
%
% For peerreview papers, this IEEEtran command inserts a page break and
% creates the second title. It will be ignored for other modes.
\IEEEpeerreviewmaketitle

\IEEEraisesectionheading{\section{Introduction}\label{sec:introduction}}
% Computer Society journal (but not conference!) papers do something unusual
% with the very first section heading (almost always called "Introduction").
% They place it ABOVE the main text! IEEEtran.cls does not automatically do
% this for you, but you can achieve this effect with the provided
% \IEEEraisesectionheading{} command. Note the need to keep any \label that
% is to refer to the section immediately after \section in the above as
% \IEEEraisesectionheading puts \section within a raised box.

% The very first letter is a 2 line initial drop letter followed
% by the rest of the first word in caps (small caps for compsoc).
% 
% form to use if the first word consists of a single letter:
% \IEEEPARstart{A}{demo} file is ....
% 
% form to use if you need the single drop letter followed by
% normal text (unknown if ever used by the IEEE):
% \IEEEPARstart{A}{}demo file is ....
% 
% Some journals put the first two words in caps:
% \IEEEPARstart{T}{his demo} file is ....
% 
% Here we have the typical use of a "T" for an initial drop letter
% and "HIS" in caps to complete the first word.

%\jmr{R pour retrait}
%\jms{S pour suggestion}
%\jmc{C pour commentaires}
%\mjm{MJM: Commentaires}

\begin{figure}[!thb]
\centering
\includegraphics[width=0.85\columnwidth]{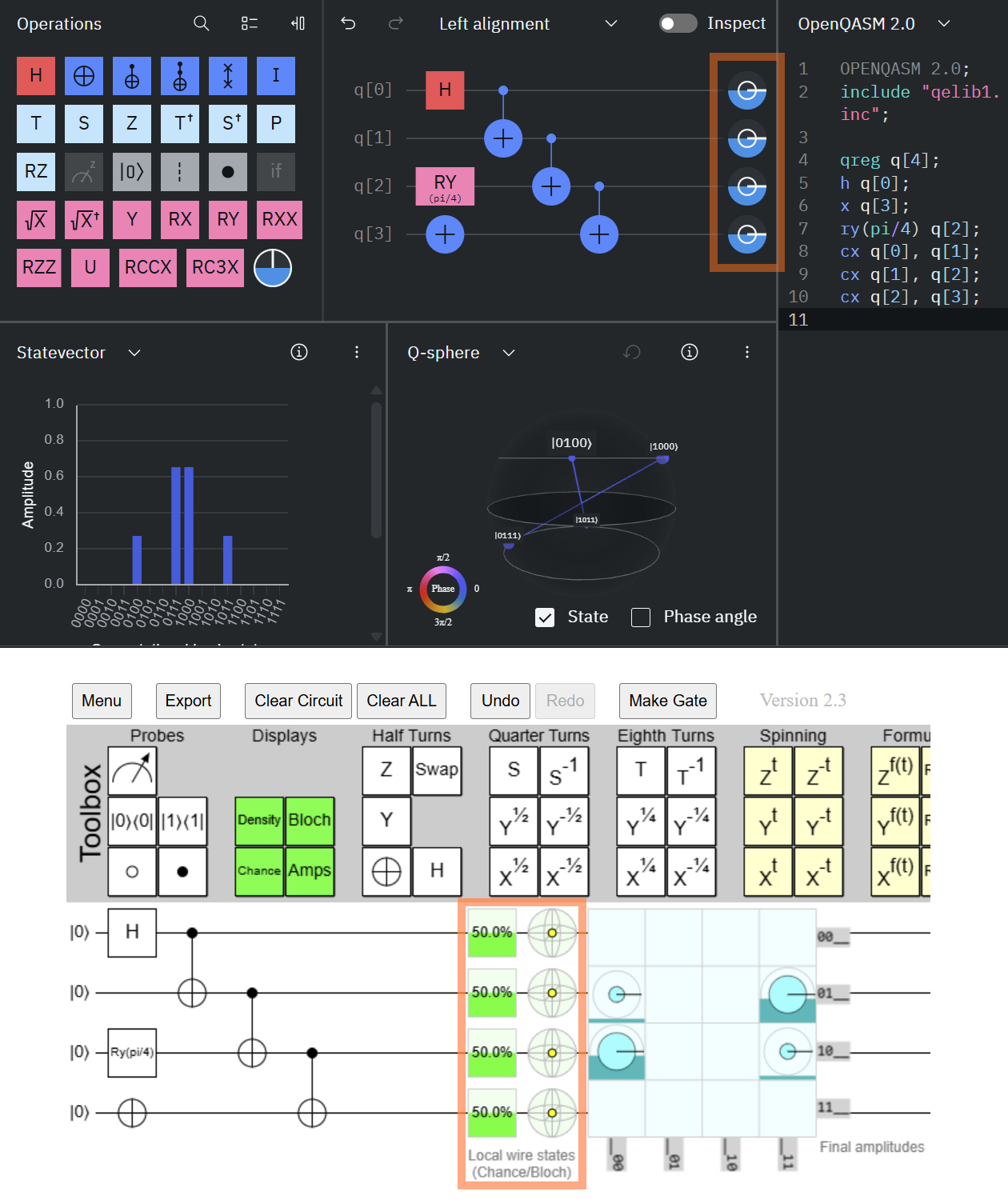}
\caption{IBM Quantum Composer (top) \cite{ibm2023composer} and Quirk (bottom) \cite{gidney2020quirk} showing the same circuit.
In both, the reduced state of the qubits is shown to the right of the circuit diagram,
in the area enclosed by the orange rectangle.
Inside this orange rectangle,
IBM's ``phase disks'' show the phase (white tick mark), probability (blue area),
and purity (the diameter of the white circle is half of the disk, indicating a purity of 0.5),
whereas Quirk's widgets show the probability (green area) and Bloch sphere coordinates
(a yellow dot in the center of each sphere, because the reduced states are maximally mixed, meaning a purity of 0.5).
}
\label{fig:ibmAndQuirk}
\end{figure}

\IEEEPARstart{G}{raphical} simulators for quantum circuits
often show a circuit diagram with small visual summaries of the reduced state (also called local state) of each qubit.
These visual summaries might show the measurement probability for the $\ket{1}$ state,
the qubit's phase or coordinates in the Bloch sphere,
and the purity in the reduced state\footnote{The reduced state of a qubit is obtained by tracing out all other qubits from the density matrix.
Sections 6.3 and 6.5 in \cite{mcguffin2025tutorial} explain how to compute these and related statistics.} (Figure~\ref{fig:ibmAndQuirk}).
Although this information is useful,
it can also be limiting or confusing for beginners:
for example, just showing information about single-qubit reduced states can make it difficult
to know which qubits are correlated or entangled,
and a qubit's phase does not always change as might be expected in response to a gate.
More fundamentally, in a circuit on $n$ qubits,
visualizing the reduced state (or any fixed amount of information) for each qubit
only shows $O(n)$ information,
which {\em cannot} reveal a complete picture of the circuit's quantum state,
since the state vector comprises $O(2^n)$ degrees of freedom.
% This limitation holds more generally for any visual approach that only shows a fixed amount of information per qubit.

%\begin{figure*}[tb]
%\begin{figure}[!thb]
\begin{figure*}[!t]
 \centering
  \ifthenelse{\boolean{UseHighResolutionImages}}{
    \includegraphics[width=1.99\columnwidth]{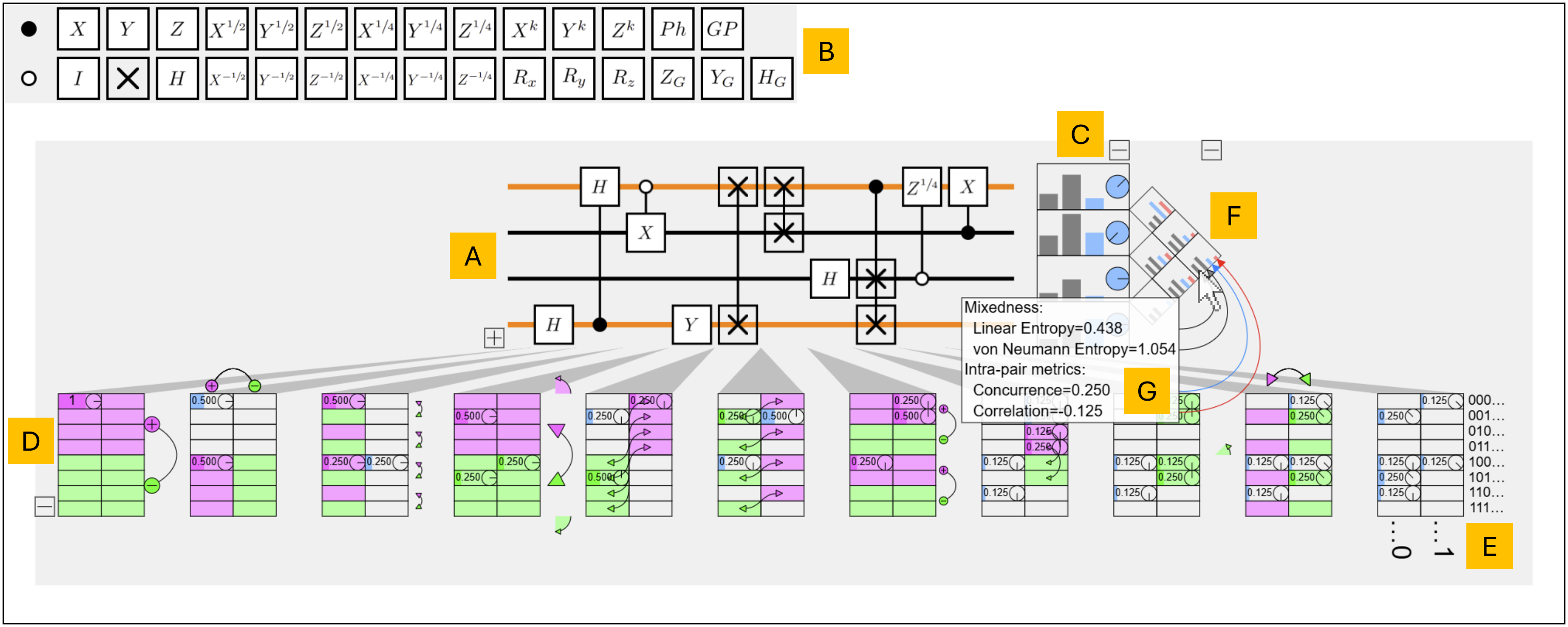}
  }{
    \includegraphics[width=1.99\columnwidth]{fig/01-teaser-2-withCursorAndAnnotations.png}
  } % ifthenelse
  \caption{
     MuqcsCraft's visual interface. %  MuqcsCraft with an example circuit, chosen to exhibit a variety of features.
     A: The circuit diagram. Black and white dots correspond to control and anticontrol qubits.
     B: A toolbar from which gates can be dragged onto the circuit diagram.
     C: The reduced state of each qubit at the output of the circuit:
     linear entropy and von Neumann entropy (shown with grey bars),
     probability of measuring 1 (blue bar),
     phase (blue disc).
     D: The state vector, layer-by-layer, shown below the circuit diagram.
     For each layer of the circuit, the 16 amplitudes of the state vector are wrapped into 8$\times$2 cells.
     Difference highlighting indicates how the state vector changes under the effect of each gate,
     using the purple and green colors, arrows, and symbols.
     E: A key to the bitstrings for the base states. For example, the lower left cell in each 8$\times$2 state vector
     is 1110, where the left-most bit corresponds to the bottom wire in the circuit.
     F: The triangular half-matrix shows one cell for each pair of qubits.
     For example, the mouse cursor is over the right-most cell,
     corresponding to the top and bottom wires which are highlighted in orange.
     G: A tooltip with curved callout arrows gives details about the contents of the cell under the cursor.
  }
 \label{fig:teaser}
\end{figure*}

As an alternative approach, we propose visualizing all $2^n$ complex amplitudes of the entire state vector,
as they evolve layer-by-layer.
This is only feasible for small circuits (roughly $n \leq 8$, or 256 amplitudes), however,
circuits of this size are useful for educational purposes and for developing intuition.
We show that a simple set of {\em visual primitives} (colors, arrows, and symbols)
is sufficient to visually augment % mark / decorate / annotate 
the state vector to elucidate the effect of a core subset of gates (Figure~\ref{fig:teaser}),
namely
Hadamard, $X$, $Y$, $Z$, $S=\sqrt{Z}$, $T=\sqrt[4]{Z}$, $Z^k$, ${\rm Phase}(\theta)$, ${\rm GlobalPhase}(\theta)$,
and SWAP gates\footnote{${\rm Phase}(\theta)$ is equivalent to $Z^{\theta/\pi}$. GlobalPhase is abbreviated as GP in the screenshots of our software. A complete list of matrix definitions of the gates is available at \url{https://github.com/MJMcGuffin/muqcs.js?tab=readme-ov-file\#matrix-definitions}},
where each gate may have any combination of control and anticontrol qubits.
We call this use of colors, arrows, and symbols {\em \svdh},
because of the way they illustrate the evolution of the state vector (Figure~\ref{fig:teaser}~D).
Other gates not in the core subset (such as $X^k$ and $Y^k$) can be replaced with equivalent sequences of gates from the subset,
enabling the effect of all single-qubit gates to be visualized with difference highlighting,
establishing a kind of {\em \vuni} which we define later.
Thus, a small set of visual primitives makes quantum state dynamics more comprehensible.

We also propose a second novel visualization called the {\em half-matrix}
which shows information about every pair of qubits (Figure~\ref{fig:teaser}~F).
Our implementation shows the entropy, correlation,
and concurrence \cite{coffman2000distributed}
for each pair of qubits,
and the half-matrix
can easily show other kinds of pairwise information.

We also present examples % case studies
demonstrating how \svdh\ can
make it more straightforward to design a circuit,
and how the half-matrix makes it easier to identify
relationships between pairs of qubits that are partially or fully entangled.
These tasks are much easier to perform with our techniques than with the status quo
visualizations in existing software (Figure~\ref{fig:ibmAndQuirk}).

Our software implementation, MuqcsCraft \cite{mcguffin2025muqcscraft},
allows the user to define a circuit with drag-and-drop actions.
Interactive coordinated highlighting
(where the user hovers the cursor over one element, causing corresponding elements elsewhere to highlight)
and tooltips with curved callout arrows (Figure~\ref{fig:teaser}~G) make the user interface more self-explanatory.
MuqcsCraft is open-source and web-based, available to use at \url{https://mjmcguffin.github.io/MuqcsCraft/}
without installing any software.
% A companion video\footnote{\url{https://youtu.be/y35QxRWeAd0}}
A companion video\footnote{\url{https://youtu.be/BCunU_gCXT4}}
demonstrates features of the software.

\section{Problems with Current Approaches}

% TODO see __NOTES_RELEVANT_TO_PAPER in NOTES-new_58.txt

% \jmc{Any $n$-dimensional unitary operator $U$ (or gate) can be written as $U = e^{i\theta/n} R$, where $R$ is unitary with $\det R = 1$. In the single-qubit case ($n=2$), this gives $U = e^{i\theta/2} R$; the global phase $e^{i\theta/2}$ has no effect on the Bloch vector, while the unitary $R$ (with $\det R = 1$) produces the corresponding rotation of the Bloch sphere.
% }

\begin{figure}[!thb]
\centering
\includegraphics[width=0.99\columnwidth]{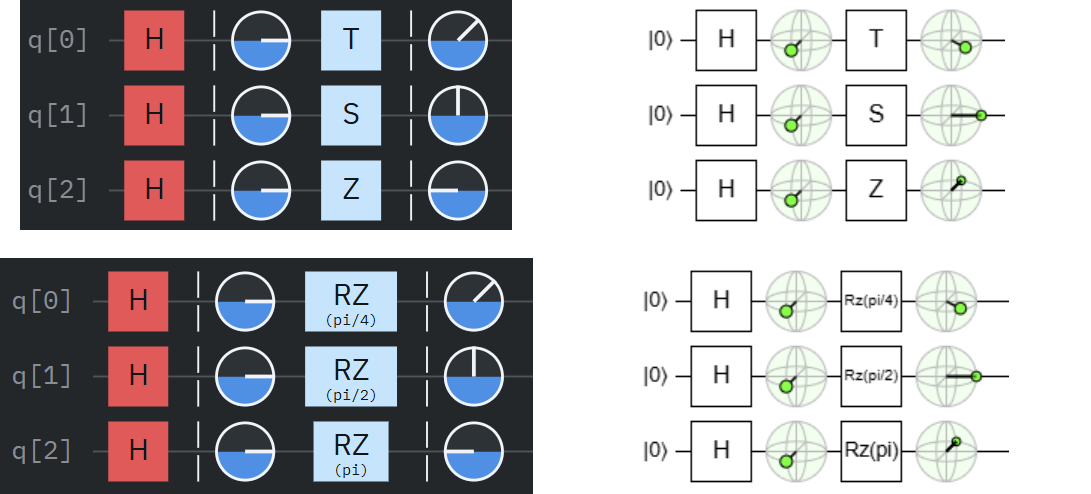}
\caption{Visual feedback in IBM Quantum Composer (left) and Quirk (right),
when gates have an easy-to-understand effect on the phase,
rotating the phase by $\pi/4$, $\pi/2$, and $\pi$ radians. % $45^{\circ}$, $90^{\circ}$, and $180^{\circ}$.
}
\label{fig:simpleRotation}
\end{figure}

\begin{figure}[!thb]
\centering
\includegraphics[width=0.99\columnwidth]{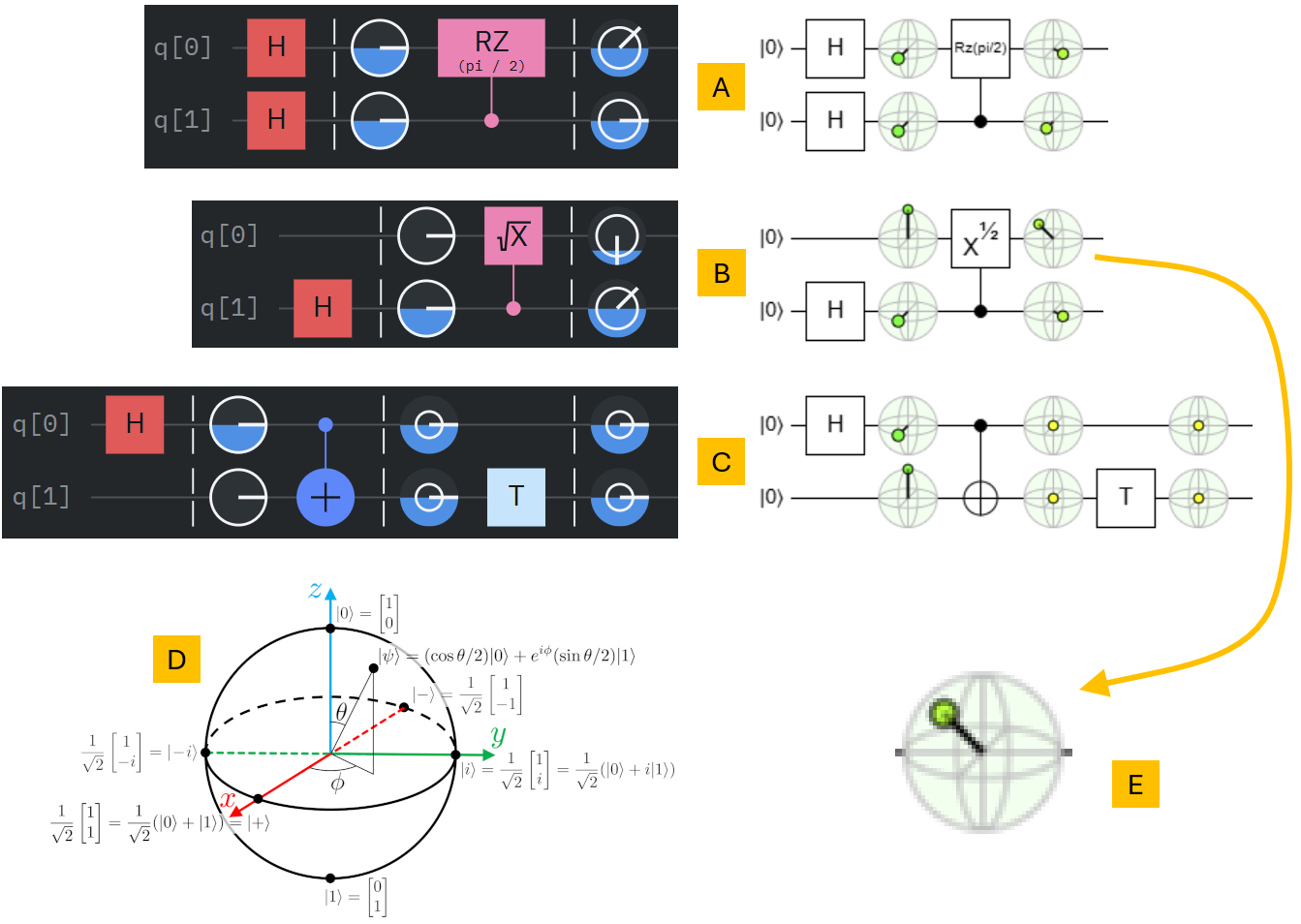}
\caption{Examples where gates have a complicated effect on qubit phase,
showing the visual feedback of IBM (left) and Quirk (right).
A: the RZ gate has an angle of $\pi/2$, but the phase changes by $\pi/4$,
because of the control qubit.
B: a $\sqrt{X}$ gate normally rotates by $\pi/2$ around the $x$ axis,
and in the circuit, we do see a change of $\pi/2$ in \texttt{q[0]}'s phase, but in the negative direction. \texttt{q[1]}'s phase also changes, because of phase kickback.
C: a $T$ gate normally rotates the phase by $\pi/4$,
but in this case, we see no change in the phase.
Each of A, B, and C could be a source of confusion for a student or beginner.
D: the Bloch sphere, where $\phi$ is the phase, and axes are oriented the same way as in Quirk's visual feedback.
E: enlarged view of Quirk's visual feedback in B.
}
\label{fig:confusingRotation}
\end{figure}

The fact that every unitary gate causes a rotation
(and/or a change in global phase)
of the state vector in Hilbert space\footnote{Any $n$-dimensional unitary operator can be expressed as $e^{ i \theta / n } R$ where $R$ is a unitary rotation matrix with $\det R = 1$.},
and that every single-qubit gate causes a rotation in the Bloch sphere (Figure~\ref{fig:confusingRotation}~D),
suggests an appealingly simple mental model.
Beginners might think that,
to change the phase of a qubit,
they merely need to choose a gate with an appropriate rotation angle.
We might therefore hope that the visualizations of qubit reduced states would guide a user
in choosing appropriate gates to achieve various effects.
Indeed, in simple cases (Figure~\ref{fig:simpleRotation}),
the effect of certain gates on qubit phase is straightforward.

%\begin{figure}[!thb]
%\centering
%\includegraphics[width=0.7\columnwidth]{fig/bloch-v2.png}
%\caption{The Bloch sphere, where $\phi$ is the phase.
%}
%\label{fig:bloch}
%\end{figure}

However, Figure~\ref{fig:confusingRotation} shows how certain cases could be
sources of confusion for a student or beginner.
First, rotation angles are reduced if there is a control qubit involved.
Second, rotations around an axis other than $z$ are not visible in IBM's ``phase disks'',
and can be difficult to see in Quirk's Bloch spheres
(for example, how easily can we see that the phase in Figure~\ref{fig:confusingRotation}~E is actually $-\pi/2$?).
Third,
single-qubit gates have no effect on Bloch sphere coordinates if the reduced state is maximally mixed,
which is not clearly shown in IBM’s user interface.
% (and, related to this, in Quirk's Bloch spheres,
% it is difficult to see if a reduced state is partially mixed or not).
Fourth, phase kickback is another potential source of confusion,
complicating the mental model of how gates affect qubits.

Even for researchers with some experience,
it might sometimes be difficult to determine which gate to use for a given purpose.
In the next section, we propose showing the state vector, and show how it can be visualized with {\em difference highlighting},
to make clearer how and why it changes from one layer of the circuit to the next.
We have found such visual feedback more helpful for designing circuits than showing qubit reduced states.

Single qubit reduced states can also make it difficult to understand a circuit's operation.
For example, Figure~\ref{fig:ibmAndQuirk}
shows a circuit similar to one that would generate a GHZ state on four qubits, except that the bottom qubit
(which is left-most in bitstrings) is inverted,
and the RY($\pi/4$) gate perturbs things.
The state vector visualizations in both user interfaces in Figure~\ref{fig:ibmAndQuirk}
indicate that the base states 0111 and 1000 are more likely than 0100 or 1011,
with all other base states having probability zero.
This implies that (1) the two top qubits (which are right-most in the bit strings)
are perfectly correlated (i.e., they always yield the same value when measured
in the computational basis);
(2) the two bottom qubits are perfectly inversely correlated;
(3) the two middle qubits are only partially correlated
and perhaps only partially entangled.
However, this is not indicated at all in the single qubit reduced states
which all indicate a purity of 0.5 in IBM's interface
(or, equivalently, maximal mixedness in Quirk's interface).
The {\em half-matrix}, presented in Section~\ref{sec:HalfMatrix},
is an alternative visual aid for analyzing circuits
that makes it easier to understand such circuits.

\section{Visualizing the State Vector}

\subsection{Basic Display Options and Features}

\begin{figure}[!thb]
\centering
\includegraphics[width=0.99\columnwidth]{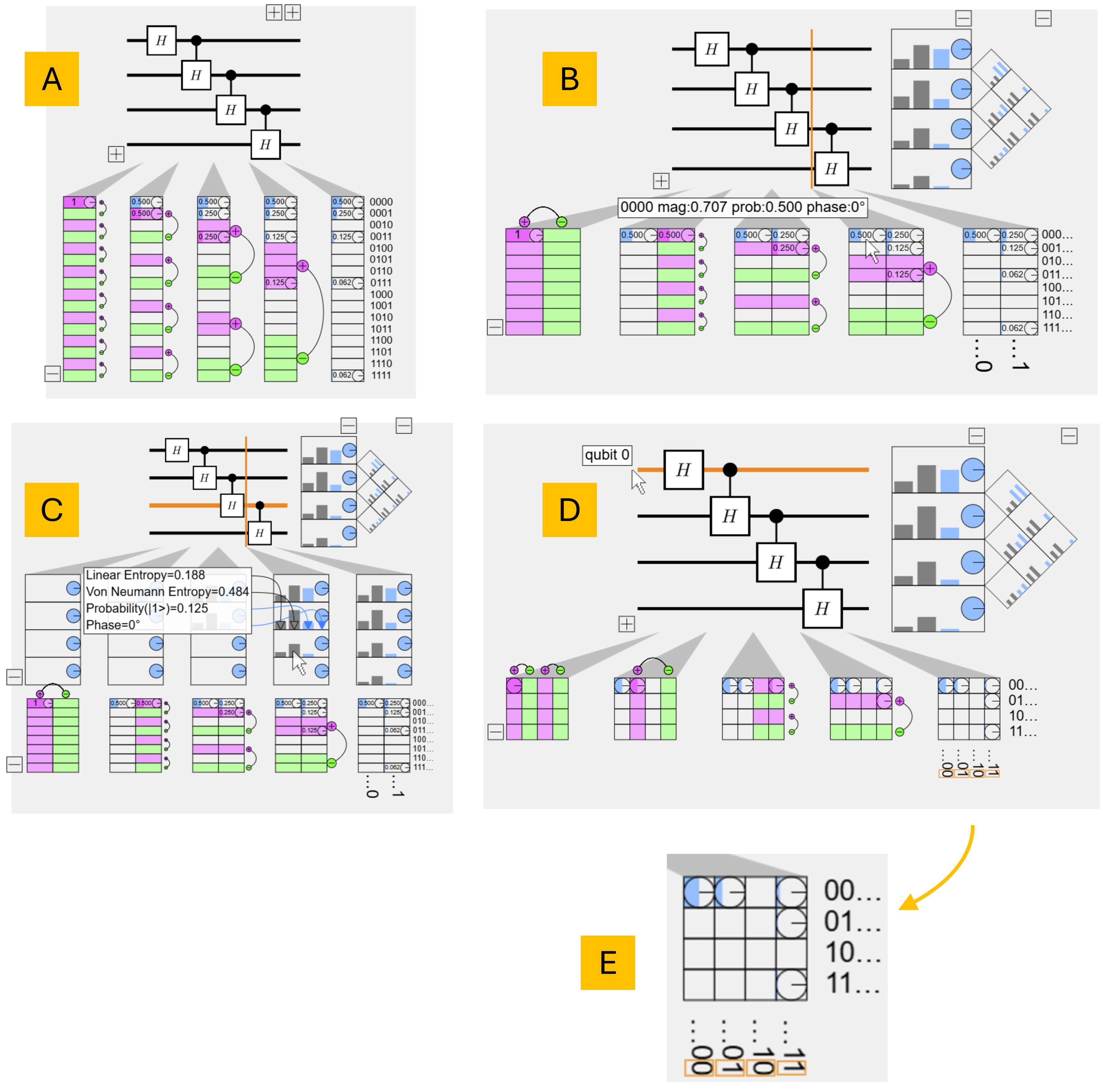}
\caption{The same circuit on 4 qubits is shown with different display options.  The state vector for each layer has 16 amplitudes.
A: each layer's state vector is shown
as a 16$\times$1 column vector.
B: each state vector is wrapped into
an 8$\times$2 arrangement.
C: the single-qubit reduced states are
also show for each layer.
D: the amplitudes of the state vector are
shown as squares,
with blue bars behind each disc instead of to the left of each disc.
E: enlarged view showing bits highlighted in orange
corresponding to the top wire under the mouse cursor in D.
}
\label{fig:displayOptions1}
\end{figure}

\begin{figure}[!thb]
\centering
\includegraphics[width=0.7\columnwidth]{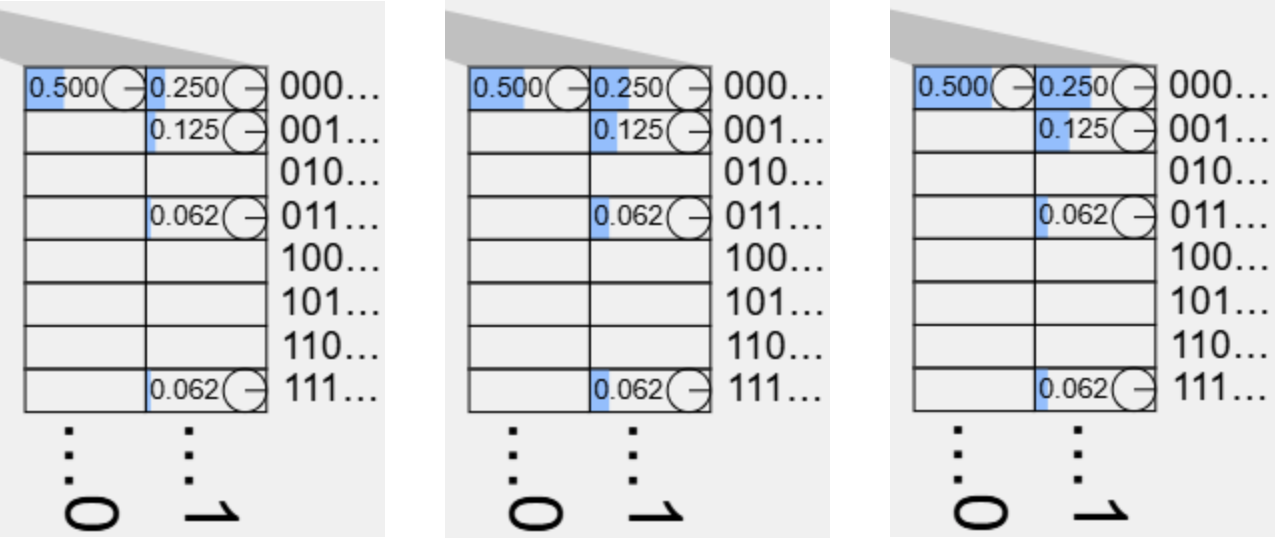}
\caption{The blue bars shown for each amplitude can have
length proportional to probability (left),
or proportional to magnitude of amplitude (middle),
or be an appropriately scaled linear function of the logarithm of
probability (right)
to make differences between small values easier to see.
}
\label{fig:displayOptions2}
\end{figure}

Figures~\ref{fig:teaser}, \ref{fig:displayOptions1} and \ref{fig:displayOptions2}
show examples of state vectors whose amplitudes are shown using horizontal blue bars
(whose horizontal length is related to the probability or magnitude, depending on the user's choice)
and discs with a tick mark (to show phase).
We now explain how the visual design of these amplitudes is based on principles from
information visualization.

% https://www.michaelmcguffin.com/MuqcsCraft/?circuit={%22cols%22:[[%22H%22],[%22%E2%80%A2%22,%22H%22],[1,%22%E2%80%A2%22,%22H%22],[1,1,%22%E2%80%A2%22,%22H%22]]}
% entire circuit:
%   no wrapping
%   with wrapping
%   with wrapping and squares
% just one state vector, wrapped with rects: each of the length functions

Probability or magnitude is shown using a bar length rather than, say, a shade of a color, because perceptual studies have established a hierarchy of visual channels (chapter 5 in \cite{munzner2014book}) that ranks length as one of the best ways to show a quantitative variable.
The bars have text labels on top of them to indicate the numerical value of the probability.  Thus, bar lengths allow a reader to quickly find large and small values, whereas the numerical text provides more precise information.
The bars are oriented horizontally, rather than vertically, to make better use of the spatial resolution made available by the text.
When an amplitude is zero, no disc is shown, to make it clearer that the amplitude is indeed zero and to make it more visually distinct from non-zero amplitudes.
Phases are shown using the {\em orientation} of the disc's tick mark, rather than, say, the length of another bar, to reflect the fact that angles wrap around at $2\pi$ back to zero.

Each state vector can be wrapped into multiple columns to save space vertically (most of the state vectors in Figures~\ref{fig:teaser}, \ref{fig:displayOptions1} and \ref{fig:displayOptions2} are 8$\times$2).
Amplitudes can also be compressed into squares
(Figure~\ref{fig:displayOptions1}~D and E)
to save even more space.

In many circuits, the magnitudes of the resulting amplitudes
are very small and
difficult to distinguish.
We implemented different definitions of bar length (Figure~\ref{fig:displayOptions2})
to allow increasing the visible difference between small values,
making it easier to see which are bigger or smaller.

Hovering the mouse cursor over different parts of the user interface
causes tooltips to appear
(Figure~\ref{fig:teaser}~G, Figure~\ref{fig:displayOptions1}~B and C)
to make the user interface more self-revealing and provide more precise information.
When appropriate, tooltips have curved callout arrows connecting them to corresponding
visual elements (Figures~\ref{fig:teaser}~G, \ref{fig:displayOptions1}~C, \ref{fig:halfMatrix}, \ref{fig:caseStudyW4}).

We also implemented multiple kinds of interactive coordinated highlighting \cite{tan2023}.
Hovering over a state vector causes a vertical orange line segment to appear
in the circuit between the corresponding layers
(Figure~\ref{fig:displayOptions1}~B).
Hovering over reduced states causes the same vertical orange line segment to appear,
and also highlights the corresponding wire in orange
(Figure~\ref{fig:displayOptions1}~C).
Hovering over a wire
causes the corresponding bits in the bitstrings to be highlighted
(Figure~\ref{fig:displayOptions1}~D and E;
% in D, the cursor is near the top wire which is highlighted in orange, and in E, the last bit of the bitstrings is also highlighted in orange;
see also Figure~\ref{fig:dhl-07-H-canceling}).
Finally, hovering over a cell in the half-matrix
causes the corresponding {\em two} wires to be highlighted in orange
(Figures~\ref{fig:teaser} and \ref{fig:halfMatrix}).

\subsection{State Vector Difference Highlighting}

Consider a layer in a circuit where there is only one gate.
The following sections
consider various kinds of gates for which
we can graphically show the effect
on the state vector.

To explain the effect of the gate,
we partition the amplitudes of the state vector into {\em even} and {\em odd} subsets.
If there are $n$ qubits,
the amplitudes are identified by bitstrings of the form $b_{n-1}\ldots b_0$.
For example, if $n=3$, the bitstrings are 000, 001, ..., 111.
If the gate is on wire $j$, where $0 \leq j < n$,
then the {\em even} amplitudes are defined as those whose bitstrings have their $j$th bit equal to 0,
and similarly, the {\em odd} amplitudes have bitstrings with $j$th bit equal to 1.
As an example, for $n=3$, $j=1$, the odd amplitudes have bitstrings
0\underline{1}0, 0\underline{1}1, 1\underline{1}0, 1\underline{1}1,
where the $j$th bit is underlined.
As another example, Figure~\ref{fig:dhl-04-X}
shows the even and odd amplitudes in different colors.
We use these definitions of odd and even in the following sections.

\subsubsection{Z gates}

The effect of a $Z$ on the state vector is to rotate
the odd amplitudes.
In our \svdh,
the subset of amplitudes affected by this
is colored green.

\begin{figure}[!thb]
\centering
\includegraphics[width=0.99\columnwidth]{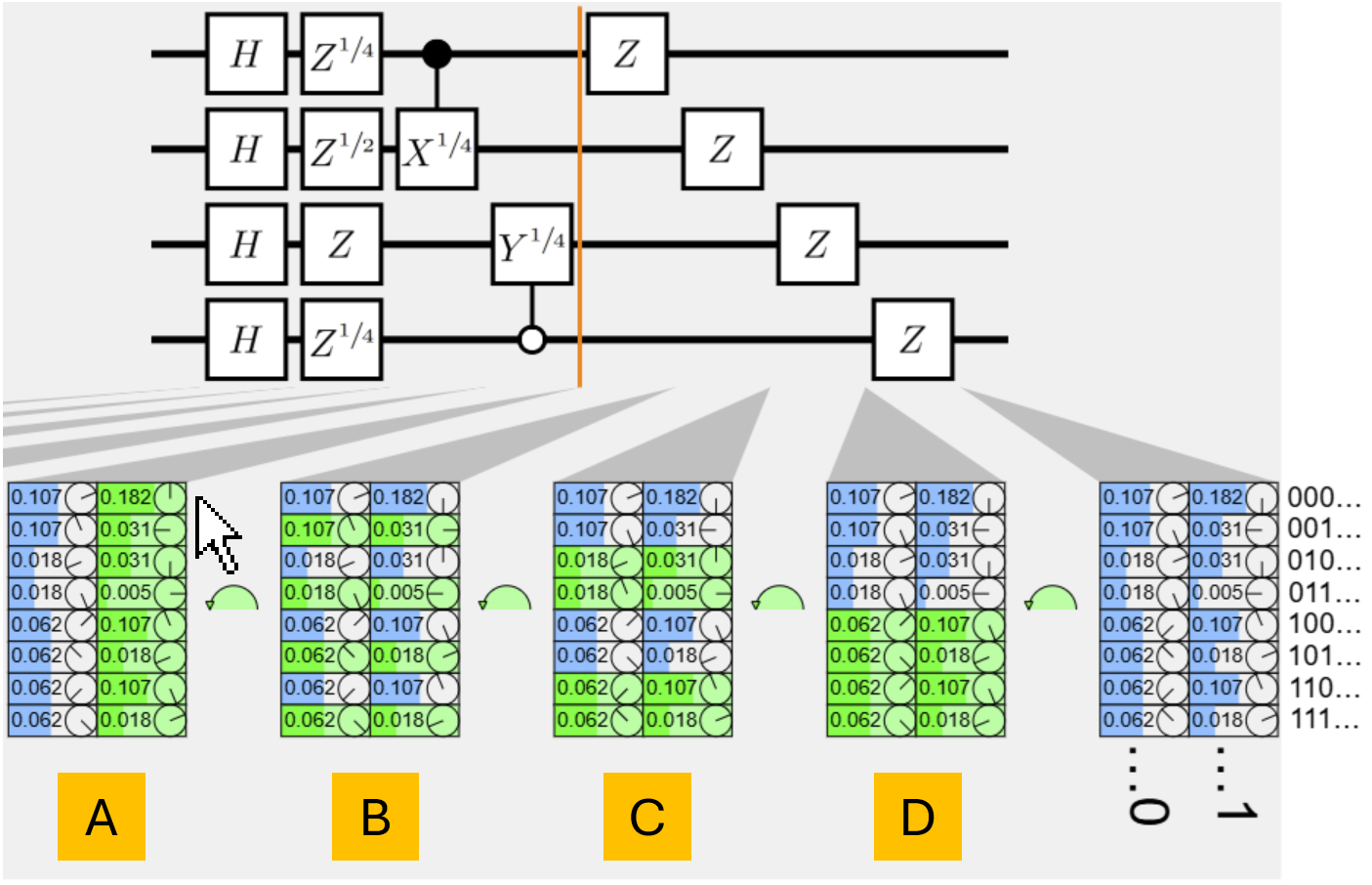}
\caption{Difference highlighting showing the effect of the
  $Z$ gates in the last 4 layers of the circuit.
  Amplitudes rotated by each $Z$ gate are colored green.
  (The first 4 layers of the circuit were designed to create
a variety of amplitudes with different magnitudes and phases,
to make it easier to see how these are
modified in the subsequent layers. These first 4 layers are reused in several subsequent figures.)
}
\label{fig:dhl-01-Z} % dhl = difference high lighting
\end{figure}

Figure~\ref{fig:dhl-01-Z} shows the difference highlighting
when the $Z$ gate is on different wires.
The green semicircular arrow to the right of each state vector
indicates that the rotation angle is always $\pi$ radians.   % $= 180^{\circ}$.

\begin{figure}[!thb]
\centering
\includegraphics[width=0.99\columnwidth]{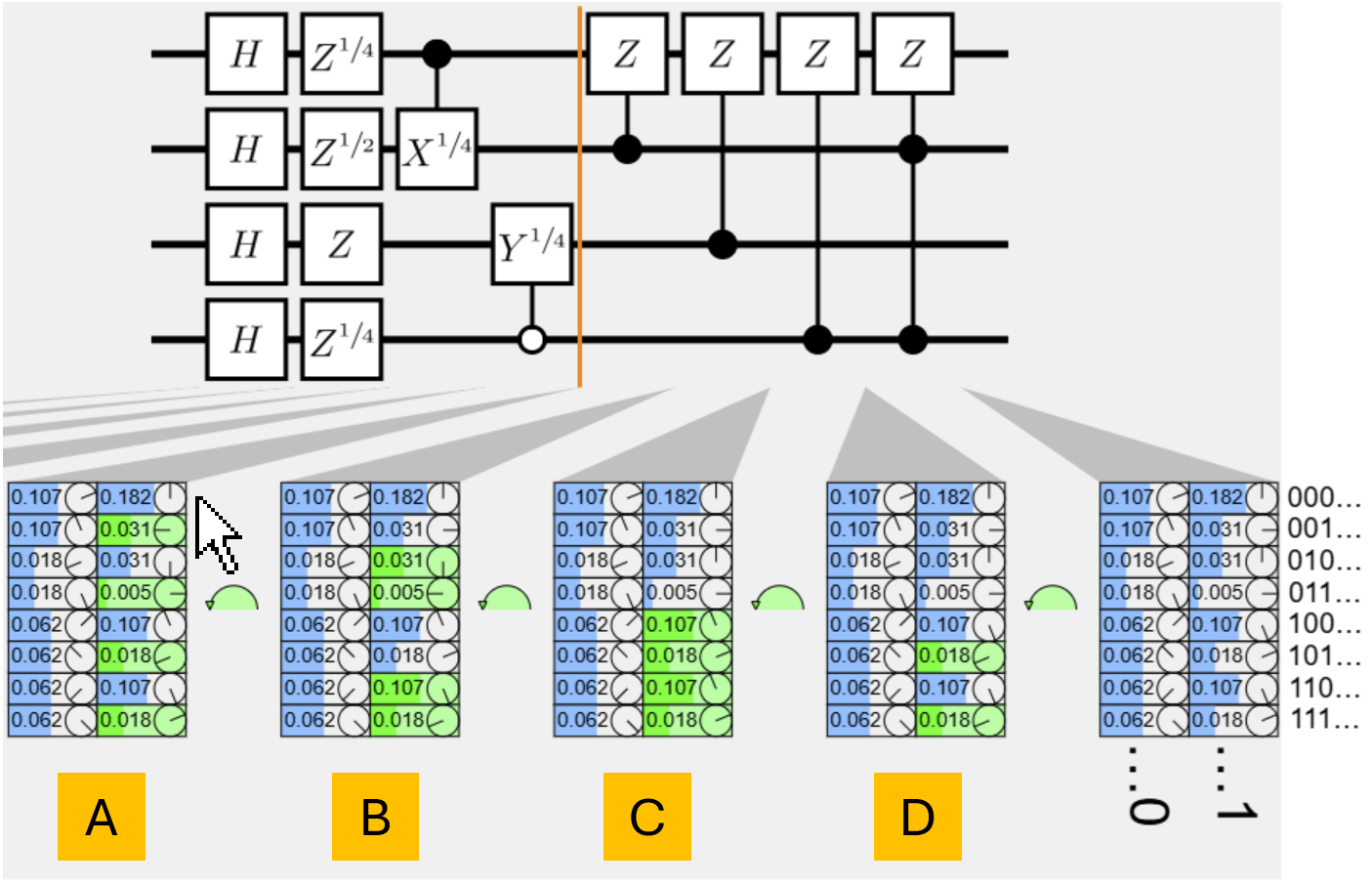}
\caption{The effect of controlled $Z$ gates is restricted by control or anticontrol qubits.
Notice that the subset of amplitudes colored green in D
is the intersection of those in A and C.
}
\label{fig:dhl-02-CZ} % dhl = difference high lighting
\end{figure}

If a $Z$ gate has one or more control (or anticontrol) qubits,
these simply limit the subset of amplitudes that are rotated.
Examples of this are shown in Figure~\ref{fig:dhl-02-CZ}.
Each additional control (or anticontrol) qubit reduces
by half the subset of amplitudes that are colored green
and that are rotated.
%   This is now redundant with the caption.
% Notice, for example, that the subset in Figure~\ref{fig:dhl-02-CZ}~D
% is the intersection of the subsets in
% Figure~\ref{fig:dhl-02-CZ}~A and \ref{fig:dhl-02-CZ}~C.

\subsubsection{S, T, Phase, and GlobalPhase gates}

\begin{figure}[!thb]
\centering
\includegraphics[width=0.99\columnwidth]{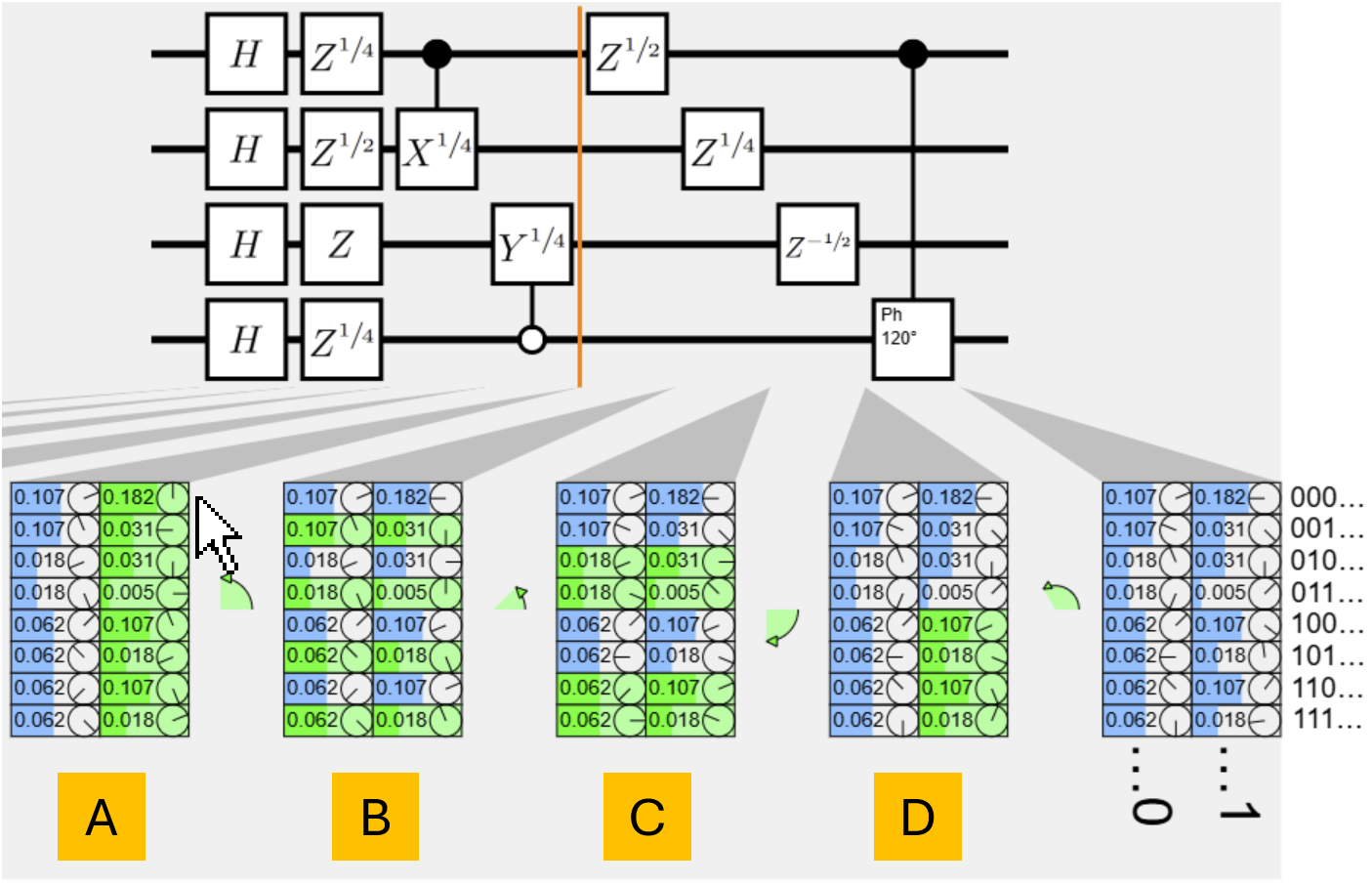}
\caption{Difference highlighting for other gates that rotate around the $z$ axis
   of the Bloch sphere.
   A: $Z^{1/2} = S$.
   B: $Z^{1/4} = T$.
   C: $Z^{-1/2} = S^{-1}$.
   D: ${\rm Phase}(120^{\circ})$ with a control qubit.
}
\label{fig:dhl-03-STPhase} % dhl = difference high lighting
\end{figure}

The difference highlighting for $Z$ gates is easily extended to
any single-qubit gates that rotate a subset of amplitudes,
even with control or anticontrol qubits.
We have done this in our software for
$S=\sqrt{Z}$ and $T=\sqrt[4]{Z}$ and their inverses,
as well as $Z^k$ and ${\rm Phase}(\theta) = Z^{\theta/\pi}$
(Figure~\ref{fig:dhl-03-STPhase}).
Our software also supports a ${\rm GlobalPhase}(\theta)$ gate,
in which case the set of amplitudes colored green is the entire state vector,
unless there are control (or anticontrol) qubits limiting the rotation to a subset.

\subsubsection{X gates}

\begin{figure}[!thb]
\centering
\includegraphics[width=0.99\columnwidth]{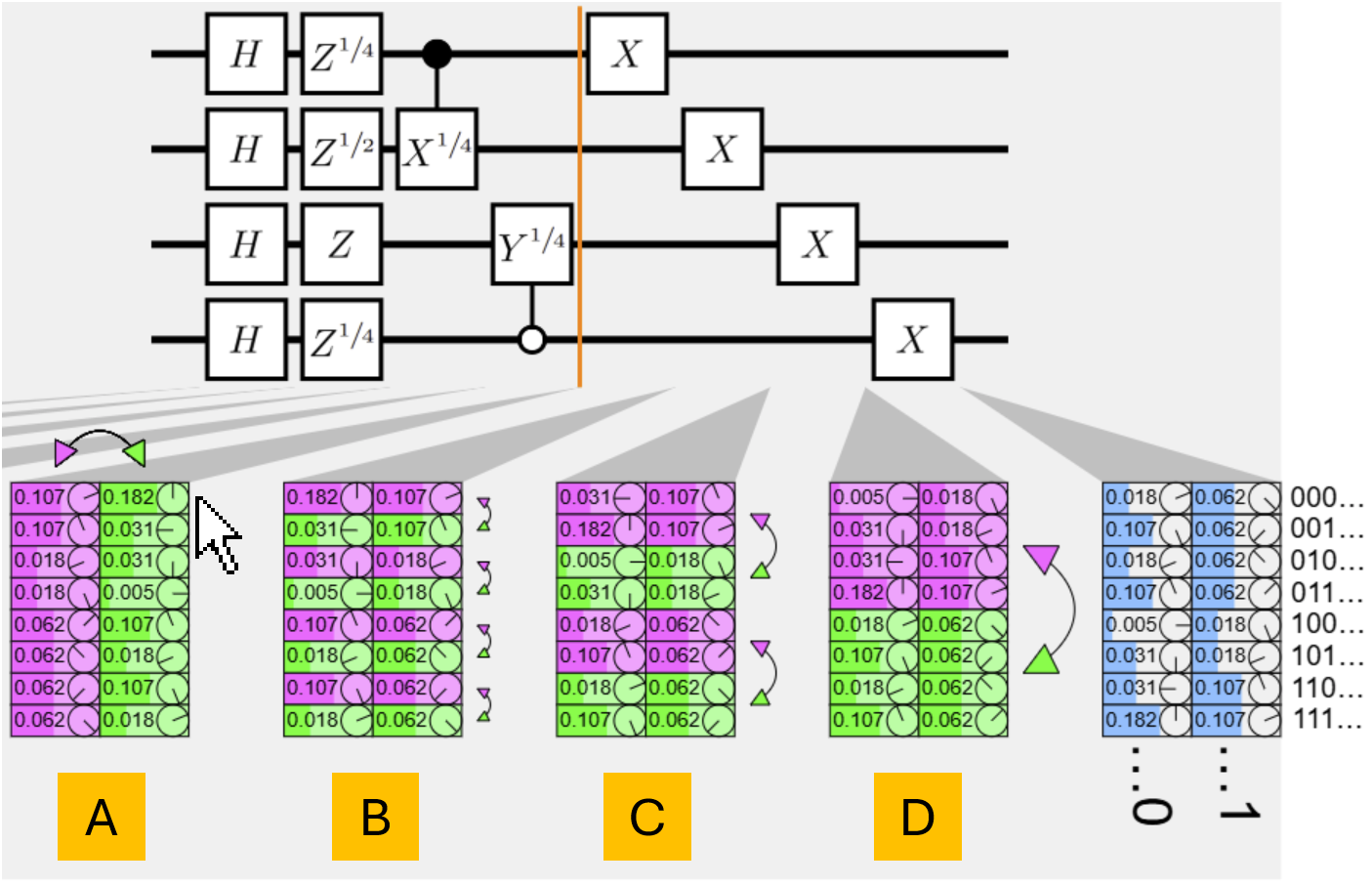}
\caption{The effect of $X$ gates on different wires.
Even amplitudes are highlighted in purple,
odd amplitudes in green.
}
\label{fig:dhl-04-X} % dhl = difference high lighting
\end{figure}

The effect of an $X$ gate is to exchange even and odd amplitudes of the state vector.
% (which is {\em not} the same as swapping qubits with a SWAP gate,
% as we see later).
To show this visually, we color the even and odd subsets
in purple and green\footnote{These colors were chosen because most forms of colorblindness involve a difficulty or inability to distinguish along the red-green dimension or the yellow-blue dimension of color space,
but purple and green differ along {\em both}
of these dimensions simultaneously.},
respectively (Figure~\ref{fig:dhl-04-X}).
The double-headed curved arrows mean that each pair of contiguous purple and green blocks
exchange places.

\begin{figure}[!thb]
\centering
\includegraphics[width=0.99\columnwidth]{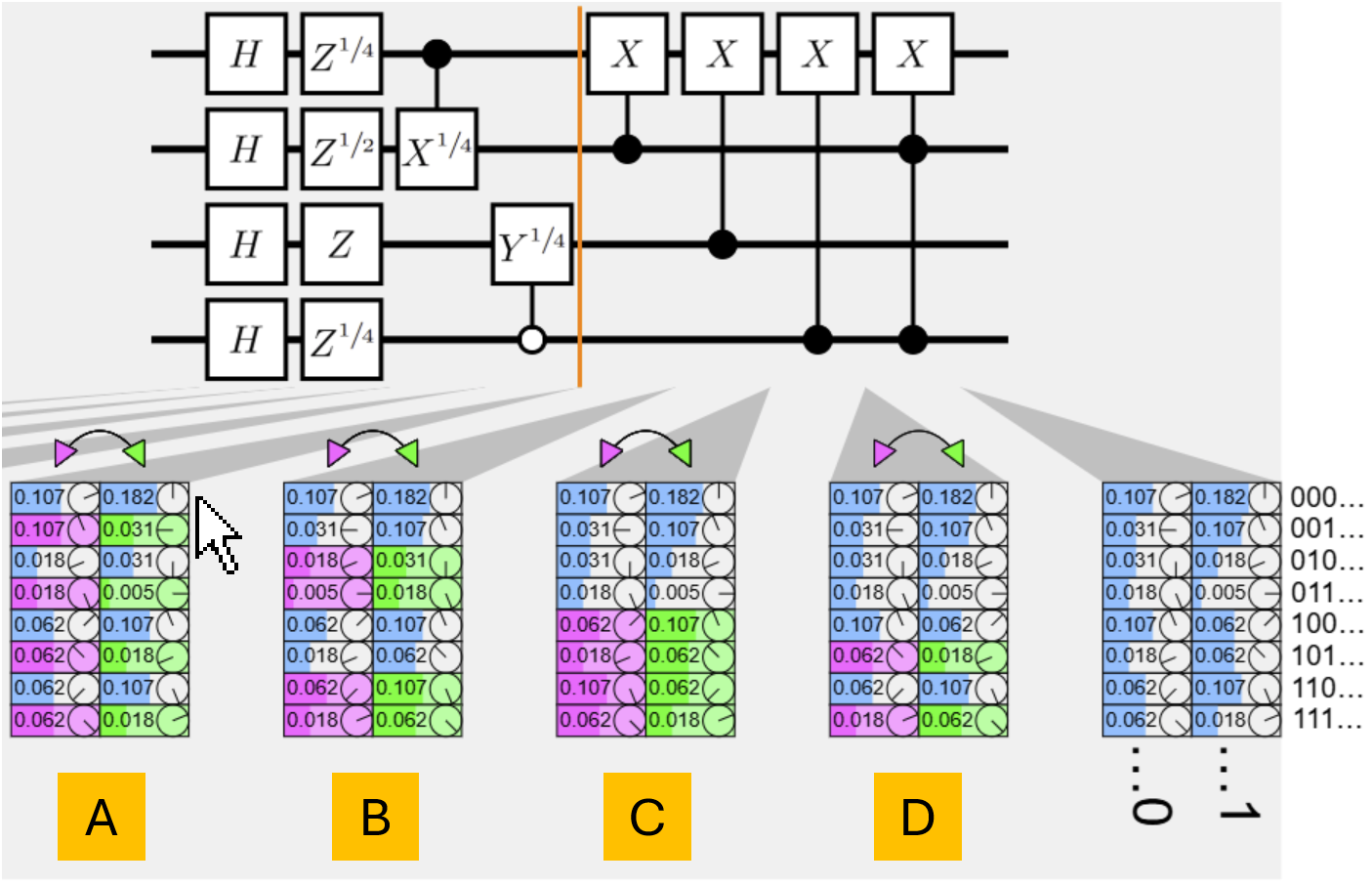}
\caption{Controlled $X$ gates. Notice that the subset colored purple and green in layer D is the
intersection of the subsets in layers A and C.
}
\label{fig:dhl-05-CX} % dhl = difference high lighting
\end{figure}

Similar to the previous examples,
adding control qubits simply restricts the effect of the $X$ gate
to a smaller subset of amplitudes (Figure~\ref{fig:dhl-05-CX}).

\subsubsection{Y gates}\label{sec:ygates}

\begin{figure}[!thb]
\centering
\includegraphics[width=0.99\columnwidth]{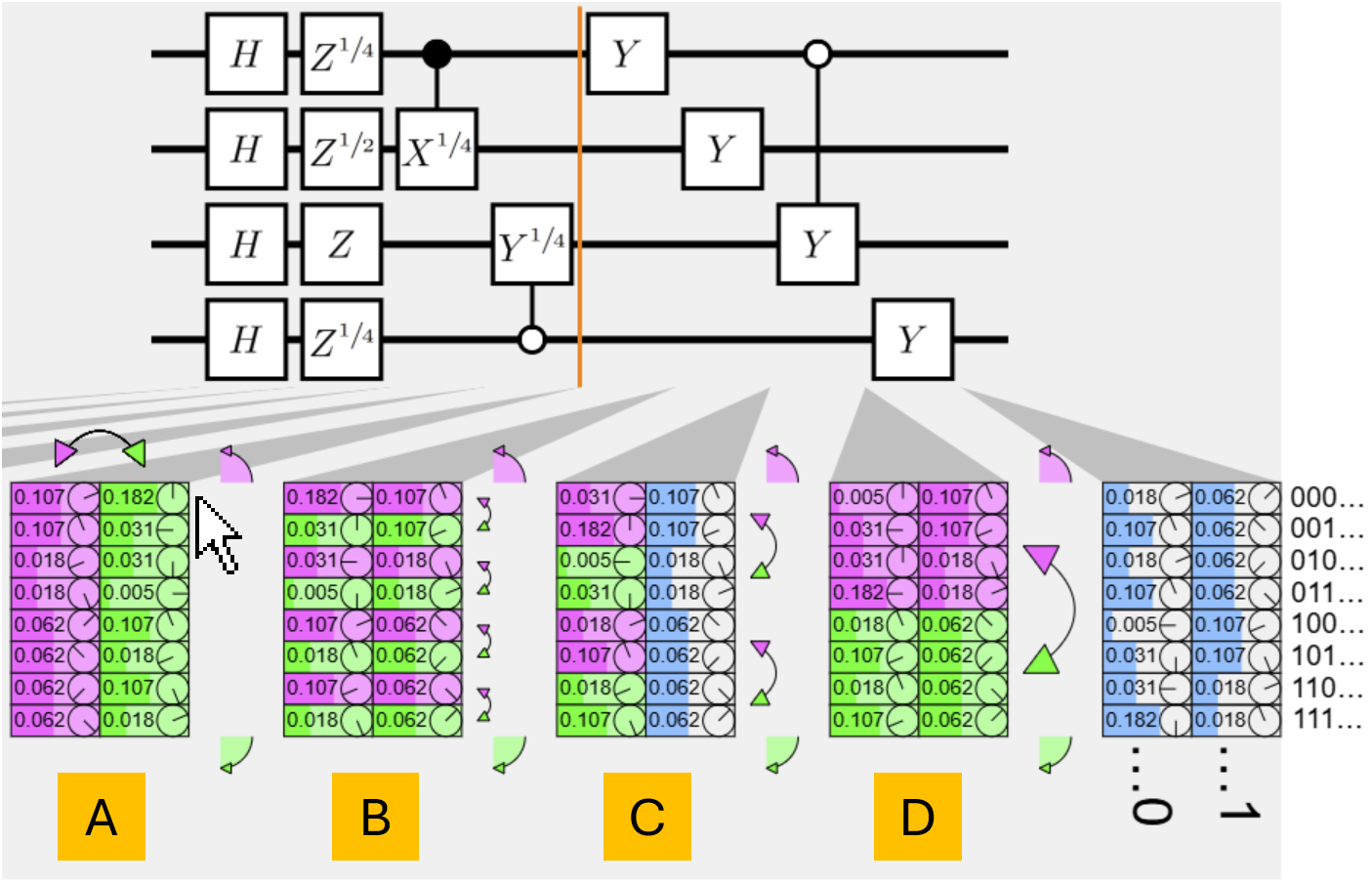}
\caption{Difference highlighting for $Y$ gates.
   For each $Y$ gate, imagine the $+\pi/2$   % $+90^{\circ}$
   purple anticlockwise rotation
   applied to the purple (even) subset,
   the $-\pi/2$ green clockwise rotation
   applied to the green (odd) subset,
   and then the purple and green subsets exchanging places
   to yield the state vector in the next layer.
}
\label{fig:dhl-06-Y} % dhl = difference high lighting
\end{figure}

Visualizing the effect of $Y$ gates is more difficult.
However, we can notice that $Y=S X S^{-1}=Z^{1/2} X Z^{-1/2}$.
Reading this right-to-left, $Z^{-1/2}$ rotates the odd amplitudes by $-\pi/2$ radians,
then $X$ exchanges the even and odd amplitudes,
and finally, $Z^{1/2}$ rotates the amplitudes that were originally even by $+\pi/2$ radians.
This is equivalent to rotating the even and odd subsets by $+\pi/2$ and $-\pi/2$, respectively,
followed by an exchange.
Thus, we can combine and extend the visual feedback already designed for the $S^{\pm 1} = Z^{\pm 1/2}$ and $X$ gates,
yielding two rotation angles (one for each colored subset)
and double-headed curved ``exchange'' arrows (Figure~\ref{fig:dhl-06-Y}).

\subsubsection{Hadamard gates}

\begin{figure}[!thb]
\centering
\includegraphics[width=0.99\columnwidth]{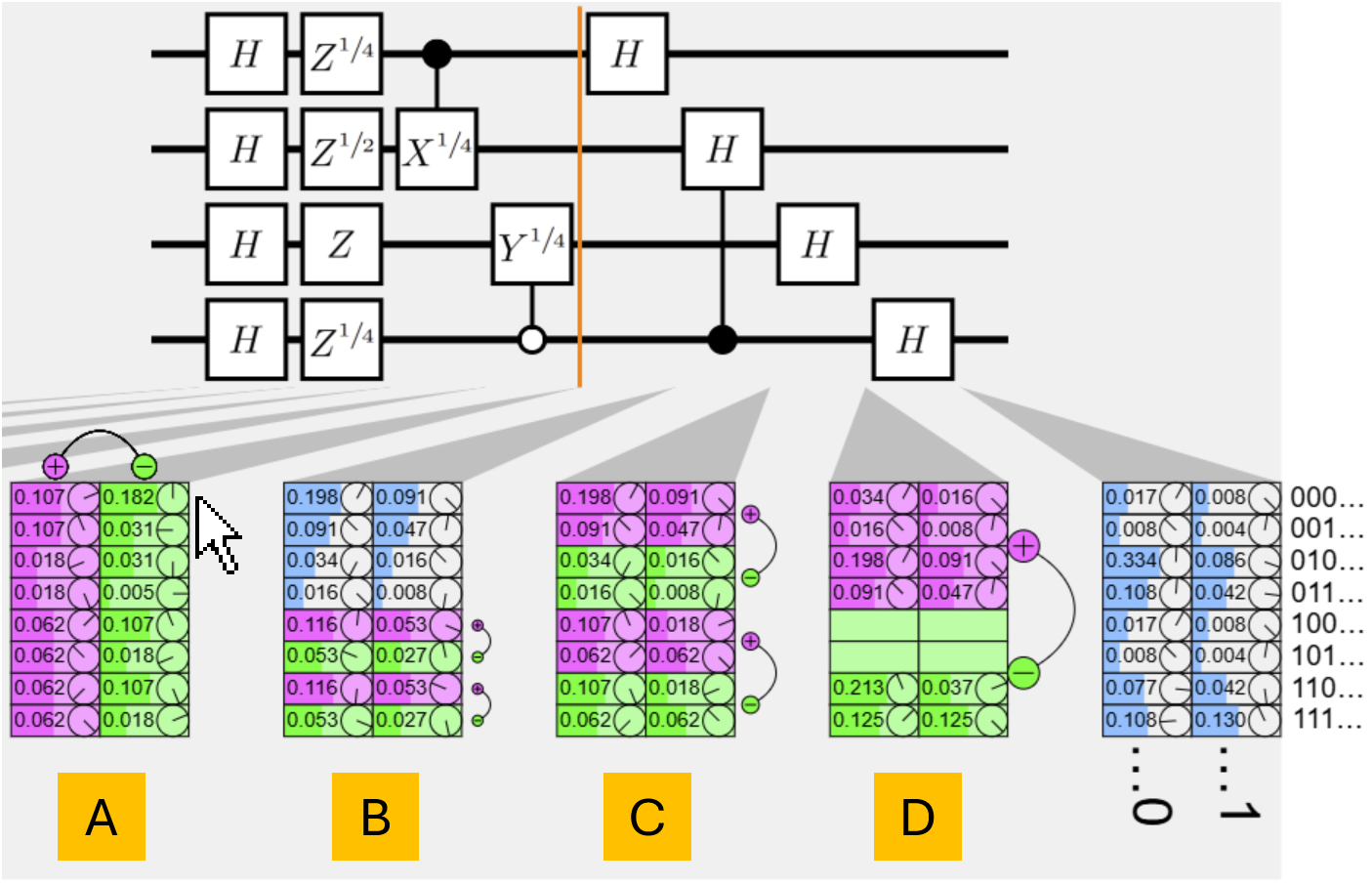}
\caption{Difference highlighting for Hadamard gates.
  For each $H$ gate,
  imagine the purple and green subsets
  being added ($\oplus$) and subtracted ($\ominus$),
  with the results replacing the purple and green subsets in the next layer,
  respectively, with an implicit scaling factor.
}
\label{fig:dhl-07-H} % dhl = difference high lighting
\end{figure}

\begin{figure}[!thb]
\centering
\includegraphics[width=0.99\columnwidth]{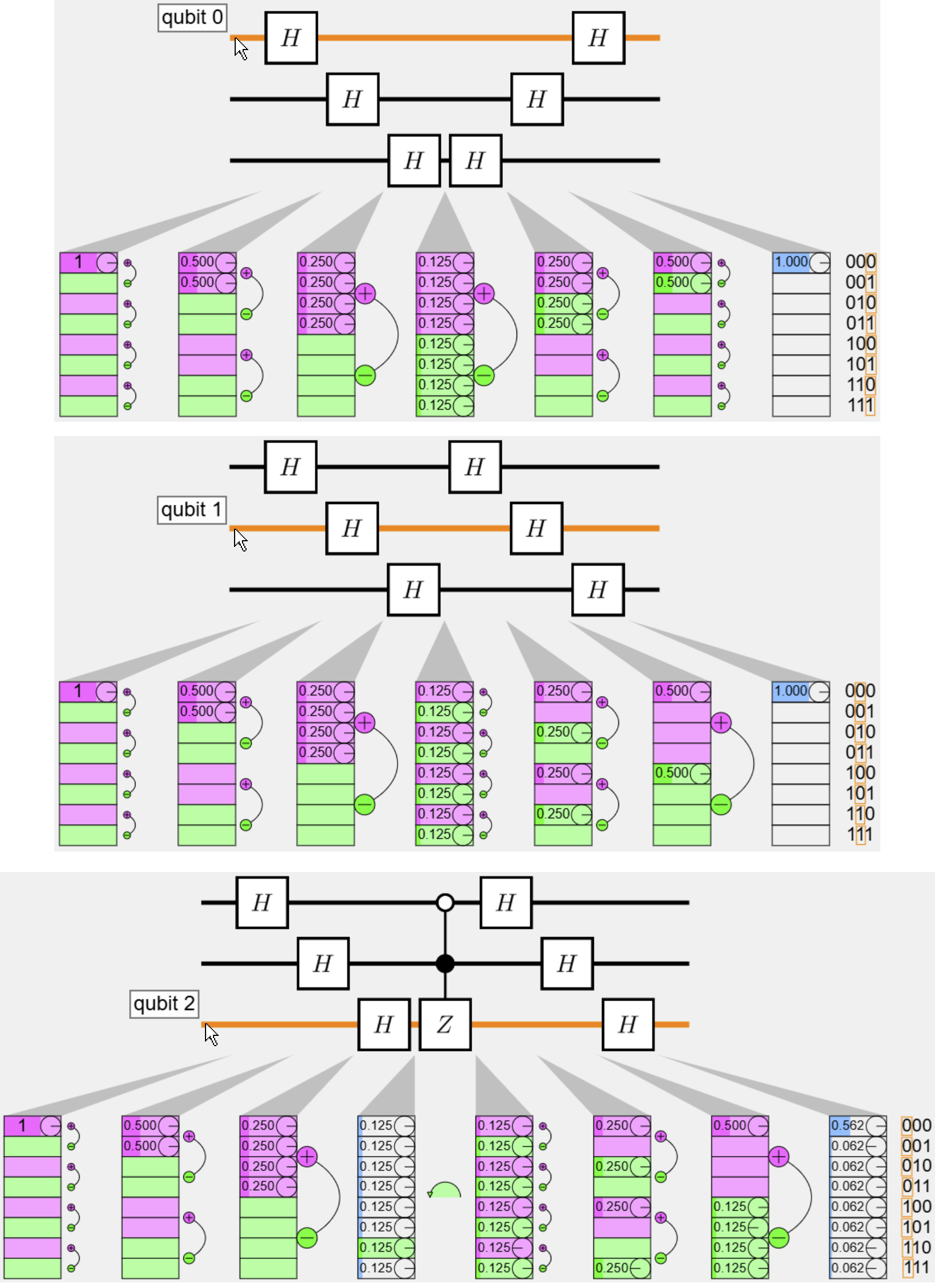}
\caption{In each circuit,
  in the first three layers, the $H$ gates
  split and distribute the nonzero amplitudes
  from purple to green amplitudes.
  In the last three layers, the subtractions cancel out
  the green amplitudes.
  Each screenshot shows a different wire and bitstring bit highlighted in orange, as a reminder of their association.
  Also, as with all screenshots throughout the article,
  the numeric labels on the state vectors show
  probabilities, not magnitudes of amplitudes.
}
\label{fig:dhl-07-H-canceling} % dhl = difference high lighting
\end{figure}

For Hadamard gates,
the concept of even and odd subsets applies again,
but now these are being added and subtracted
and scaled by $1/\sqrt{2}$.
The addition and subtraction is represented by the $\oplus$
and $\ominus$ symbols (Figure~\ref{fig:dhl-07-H}),
and the scaling factor is implicit.
The intention is to show how
multiple Hadamard gates can distribute
a single non-zero amplitude over many base states
(Figure~\ref{fig:dhl-07-H-canceling}, first three layers
of each of the circuits)
and also used to cancel amplitudes
(Figure~\ref{fig:dhl-07-H-canceling}, last three layers
of each of the circuits).

To better understand Figure~\ref{fig:dhl-07-H-canceling},
keep in mind that the numeric labels on the state vectors show
probabilities, and let $p_e$ and $p_o$ be the probabilities
for two matching base states,
where the first is even (purple)
and the second is odd (green).
If we assume that the phases of the relevant amplitudes
are oriented in the same direction,
then the probability values of $(p_e,p_o)$,
under the effect of a Hadamard gate,
will be replaced in the next layer with
$(\frac{1}{2}(\sqrt{p_e}+\sqrt{p_o})^2,\frac{1}{2}(\sqrt{p_e}-\sqrt{p_o})^2)$.
Now consider three cases.

{\em Case 1}: $p_o=0$, as in the first 3 layers of
the circuits in Figure~\ref{fig:dhl-07-H-canceling}.
In this case, each $(p_e,0)$ pair is replaced with
$(\frac{1}{2}(\sqrt{p_e}+\sqrt{0})^2,\frac{1}{2}(\sqrt{p_e}-\sqrt{0})^2) = (\frac{1}{2}p_e,\frac{1}{2}p_e)$,
spreading the probability evenly between the two base states.

{\em Case 2}: we have the same probability $p_e = p_o$ on both base states, and the relevant phases have the same orientation, as seen in the last 3 layers
of the top and middle circuits of
Figure~\ref{fig:dhl-07-H-canceling}.
Doing the calculation, we find that each $(p_e,p_e)$ pair is replaced
with $(2 p_e,0)$, concentrating the probability on the even base state.

{\em Case 3}: $p_e = p_o$, but the relevant phases are $\pi$ radians   % $180^{\circ}$
different, as seen in one case in the bottom circuit
of Figure~\ref{fig:dhl-07-H-canceling}, immediately after the $Z$ gate,
for the base states 110 and 111.
In this case, we calculate that $(p_e,p_e)$
is replaced with $(0,2 p_e)$, concentrating probability on the odd base state.

\subsubsection{SWAP gates}

To understand the effect of a SWAP gate on the state vector,
consider as an example that there are $n=4$ qubits, hence $2^n=16$ base states
represented by all the binary strings of length 4,
and we can write these strings as $b_3 b_2 b_1 b_0$ where $b_k$ is the $k$th binary bit.
Assume also that there is a SWAP gate acting on qubits $i$ and $j$, where $0 \leq i < j < n$.
To take a concrete example, assume $i=1$ and $j=3$.
Then, the amplitudes for the following pairs of base states
will be swapped:
(\underline{0}0\underline{1}0, \underline{1}0\underline{0}0),
(\underline{0}1\underline{1}1, \underline{1}1\underline{0}1),
and any pair of the form
(\underline{0}$b_2$\underline{1}$b_0$, \underline{1}$b_2$\underline{0}$b_0$),
% where $b_k$ is the $k$th binary bit,
where we have underlined bits $b_i$ and $b_j$.

\begin{figure}[!thb]
\centering
\includegraphics[width=0.99\columnwidth]{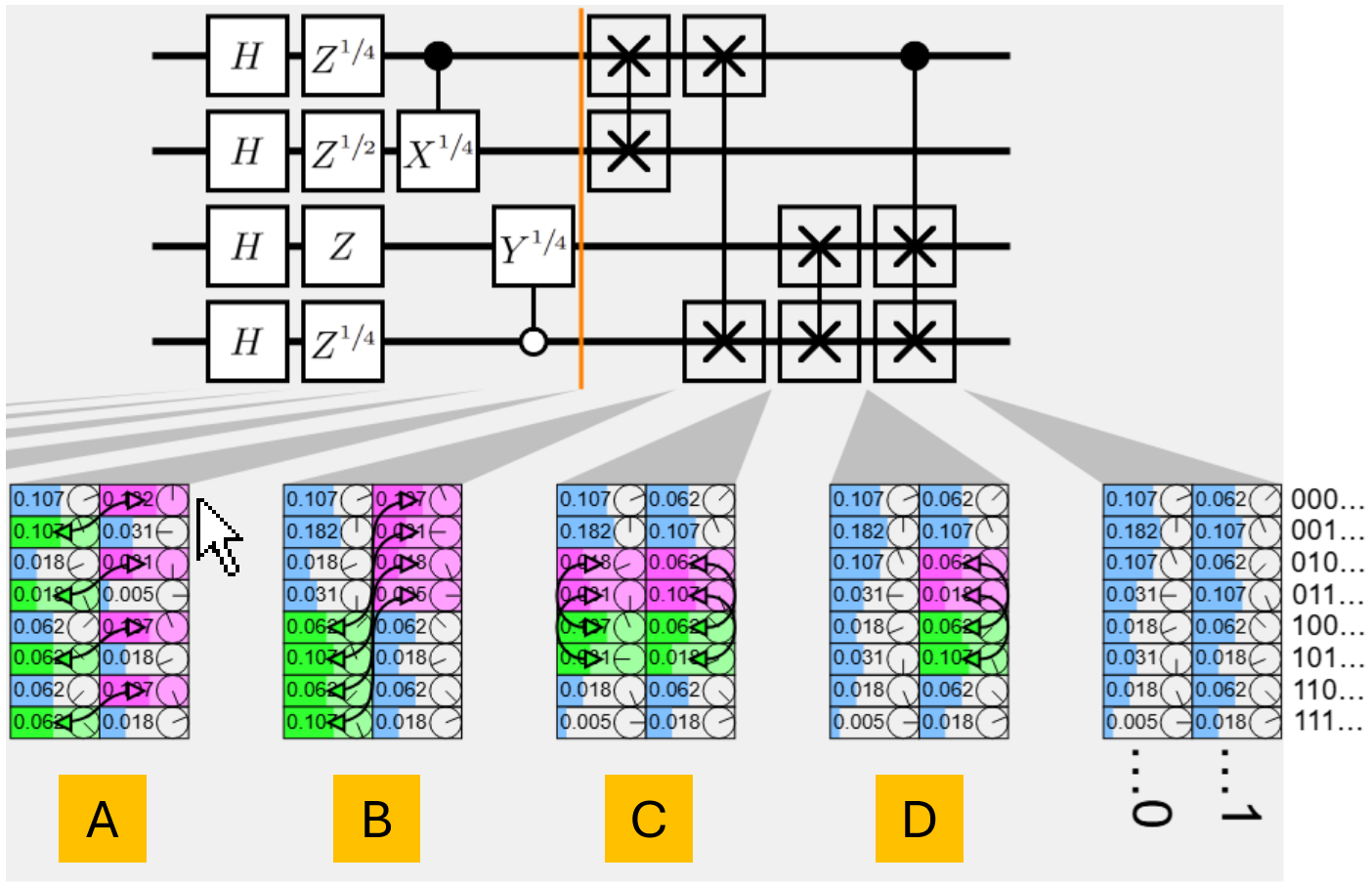}
\caption{Difference highlighting with SWAP and controlled-SWAP gates.
}
\label{fig:dhl-08-SWAP} % dhl = difference high lighting
\end{figure}

In our visualization, we use double-headed arrows
to show which amplitudes are swapped.
Assume that the state vectors
are wrapped and displayed as $2^{n-K}$ rows $\times 2^K$ columns
(in Figure~\ref{fig:dhl-08-SWAP}, $n=4$ and $K=1$).
In each layer, there are three possible cases.
If $K \leq i < j$, then each pair of amplitudes that are swapped appear in the same column (Figure~\ref{fig:dhl-08-SWAP}, layers C, D).
If $i < K \leq j$, then each pair of amplitudes that are swapped appear in diagonally located positions (Figure~\ref{fig:dhl-08-SWAP}, layers A, B).
If $i<j<K$, then each pair of amplitudes that are swapped appear in the same row (not shown).

%Let $n$ be the number of qubits in the circuit, and assume that the state vectors
%are wrapped and displayed as $2^{n-k}$ rows $\times 2^k$ columns
%(in Figure~\ref{fig:dhl-08-SWAP}, $n=4$ and $k=1$).
%Also, in a given layer, assume there is a SWAP gate acting on qubits $i$ and $j$, where $0 < i < j < n$.
%If $k<=i<j$, then each pair of amplitudes that are swapped appear in the same column (Figure~\ref{fig:dhl-08-SWAP}, layers C, D).
%If $i < k <= j$, then each pair of amplitudes that are swapped appear in diagonally located positions (Figure~\ref{fig:dhl-08-SWAP}, layers A, B).
%If $i<j<k$, then each pair of amplitudes that are swapped appear in the same row (not shown).

\subsubsection{Visual universality}

We have demonstrated how a small set of visual primitives,
namely the coloring of even and odd subsets
(in purple and green),
curved arrows for rotation angles,
and symbols for exchanging (double-headed arrows),
adding and subtracting ($\oplus$ and $\ominus$),
allow us to elucidate the effect of the following core set of gates:
Hadamard,
$X$,
$Y$,
$Z$,
$S=\sqrt{Z}$, $S^{-1}$,
$T=\sqrt[4]{Z}$, $T^{-1}$,
$Z^k$,
${\rm Phase}(\theta)$,   % = Z^{\theta/\pi}$,
${\rm GlobalPhase}(\theta)$,
and SWAP,
each with arbitrary control and anticontrol qubits.
This includes, by the way, Toffoli and Fredkin gates.
However, ideally we would support arbitrary gates.
To articulate this goal, we propose a definition:
a set of visual primitives is {\em visually universal},
with respect to a set of objects,
if any of those objects can be visualized using the primitives.
Thus, we seek visual universality for our set of primitives,
with respect to the set of single-qubit unitary gates
with arbitrary control and anticontrol qubits.

It turns out that any gate not in our core set
can be rewritten as a product of gates from the core set.
By replacing individual gates outside the core set
with sequences of core gates,
we can visualize the effects on the state vector,
at the cost of increasing the depth of the circuit.
Below we list various gates and their equivalent expansions, with their {\em cost},
defined as the number of layers required to replace each original gate.
Note that, in many applications, a designer might not
be interested in changes to global phase,
in which case GlobalPhase gates can be removed from the expansions,
saving one layer.
~ \\
%\begin{table}[h]
% \caption{Expansions of gates into core gates}
% \label{tab:expansions}
\scalebox{0.99} {
  \begin{tabular}{@{}c@{~}|l|@{~}c@{}}
    \hline
     Gate & Expansion into core gates & Cost  \\
    \hline
    $X^{1/2}$           &  $H S H$ & 3 \\ 
    $X^{-1/2}$          &  $H S^{-1} H$ & 3 \\ 
    $Y^{1/2}$           &  $H Z {\rm GlobalPhase}(\pi/4)$ & 3 \\ 
    $Y^{-1/2}$          &  $Z H {\rm GlobalPhase}(-\pi/4)$ & 3 \\ 
    $X^k$               &  $H Z^k H$ & 3 \\ 
    $Y^k$               &  $H S^{-1} H Z^k H S H$ & 7 \\ 
    ${\rm RX}(\theta)$  &  $H Z^{\theta/\pi} H {\rm GlobalPhase}(-\theta/2)$ & 4 \\
    ${\rm RY}(\theta)$  &  $H S^{-1} H Z^{\theta/\pi} H S H {\rm GlobalPhase}(-\theta/2)$ & 8 \\
    ${\rm RZ}(\theta)$  &  $Z^{\theta/\pi} {\rm GlobalPhase}(-\theta/2)$  &   2 \\
    \hline
  \end{tabular}
} % scalebox
%\end{table}
~ \\

It is also well known that {\em any} single-qubit unitary gate
can be written as a product of appropriate rotation gates,
and there exist algorithms to convert any multi-qubit gate
into a sequence of multi-controlled single-qubit gates
\cite{mottonen2004,shende2006},
establishing the visual universality that we sought.

\subsubsection{Generalized gates enabling shorter expansions}

The visual feedback shown for $Y$ gates (\S~\ref{sec:ygates}) is the
most complicated presented so far,
since it combines {\em two} rotation angles
with a double-headed arrow,
and yet the rotation angles are
fixed at $\pm \pi/2$ radians.    %  $+90^{\circ}$ and $-90^{\circ}$
This implies that we are not making
full use of the degrees of freedom available in our set of visual primitives.
We therefore propose three new gates
that allow these rotation angles to be independently varied and combined with
the double-headed arrow and the addition and subtraction symbols.
Although these new gates are not standard, their effect on the state vector is
easy to understand using our set of primitives,
and they also allow gates outside the core set to be expanded into {\em shorter} equivalent sequences of core gates.

We now define the three new gates to be added to the core set,
which we call
{\em Generalized Z}, {\em Generalized Y}, and {\em Generalized Hadamard} gates.
Each takes two angle parameters $a$, $b$:
\begin{align*}
Z_G(a,b) = X \cdot Z^{a/\pi} \cdot X \cdot Z^{b/\pi} = \left[ \begin{array}{@{}cc@{}} e^{ia} & 0  \\ 0 & e^{ib} \end{array} \right] \\[2ex]
Y_G(a,b) = Z^{a/\pi} \cdot X \cdot Z^{b/\pi} = \left[ \begin{array}{@{}cc@{}} 0 & e^{ib} \\ e^{ia} & 0 \end{array} \right] \\[2ex]
H_G(a,b) = H \cdot Z_G(a,b) = \frac{1}{\sqrt{2}}\left[ \begin{array}{@{}cc@{}} e^{ia} & e^{ib}  \\ e^{ia} & -e^{ib} \end{array} \right]
\end{align*}

% Z^{a/\pi} GlobalPhase(b) = Z_G(b,a+b)
%     RZ(\theta) = Z^{\theta/\pi} {\rm GlobalPhase}(-\theta/2) = Z_G(-\theta/2,\theta/2)
%     Z {\rm GlobalPhase}(\pi/4) = Z_G(\pi/4,5\pi/4)
% H Z^{a/\pi} = H_G(0,a)
%     H S = H_G(0,\pi/2)    
%     H S^{-1} = H_G(0,-\pi/2)
%     H Z^{\theta/\pi} = H_G(0,\theta)
% H GlobalPhase(a) = H_G(a,a)
%     H {\rm GlobalPhase}(-\theta/2) = H_G(-\theta/2,-\theta/2)

To gain intuition for how these gates work, read the definition of the $Z_G$ gate from right-to-left: first, the $Z^{b/\pi}$ rotates the odd amplitudes by angle $b$; next, the $X$ exchanges the even and odd amplitudes; next, the $Z^{a/\pi}$ rotates the amplitudes that were originally even by angle $a$; and finally, the left-most $X$ exchanges the amplitudes again.  The net effect is that the even and odd amplitudes are rotated by angles $a$ and $b$, respectively.  The $Y_G$ gate works similarly, except it leaves the even and odd amplitudes exchanged.  And the $H_G$ gate applies a Hadamard after rotating the even and odd subsets.

\begin{figure}[!thb]
\centering
\includegraphics[width=0.99\columnwidth]{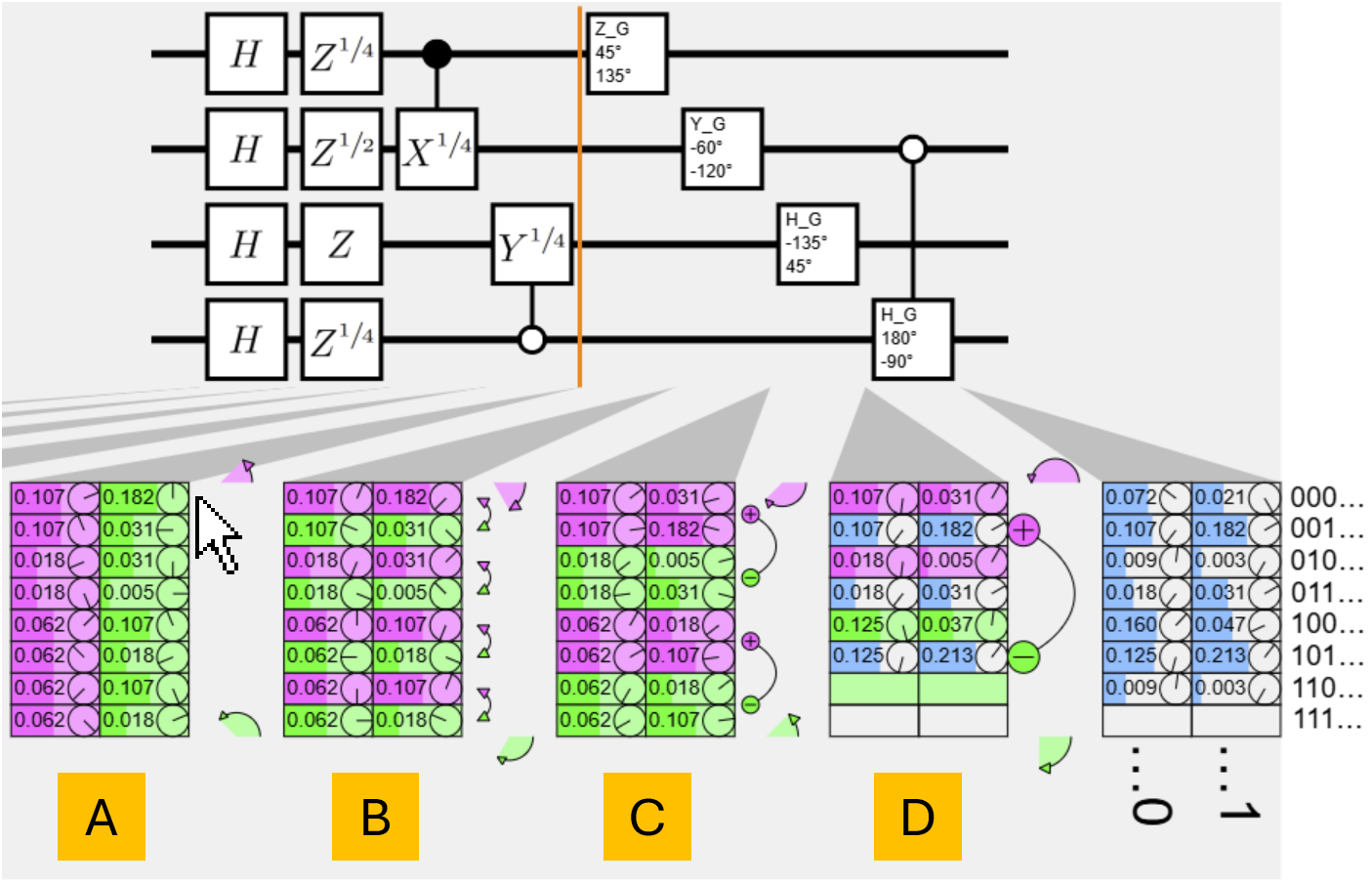}
\caption{Difference highlighting for the generalized $Z_G$, $Y_G$, $H_G$ gates.
   For each of these gates, imagine the purple rotation
   applied to the purple subset,
   and the green rotation
   applied to the green subset.
   Then, in the case of the $Y_G$ gate (layer B),
   the subsets exchange places.
   In the case of the $H_G$ gate (layers C and D),
   imagine the purple and green subsets
   being added ($\oplus$) and subtracted ($\ominus$),
   with the results replacing the purple and green subsets,
   respectively, with an implicit scaling factor.
}
\label{fig:dhl-09-ZGYGHG} % dhl = difference high lighting
\end{figure}

These generalized gates are visualized as shown in Figure~\ref{fig:dhl-09-ZGYGHG},
and they enable the following improved, shorter expansions of non-core gates:
~ \\
%\begin{table}[h]
% \caption{Expansions of gates into core gates}
% \label{tab:expansions}
\scalebox{0.89} {
  \begin{tabular}{@{}c@{~}|l|@{~}c@{}}
    \hline
     Gate & Expansion into core gates using generalized gates & Cost  \\
    \hline
    $X^{1/2}$           &  $H_G(0,\pi/2) H$ & 2 \\ 
    $X^{-1/2}$          &  $H_G(0,-\pi/2) H$ & 2 \\ 
    $Y^{1/2}$           &  $H_G(\pi/4,5\pi/4)$ & 1 \\ 
    $Y^{-1/2}$          &  $Z_G(-\pi/4,-5\pi/4) H$ & 2 \\ 
    $X^k$               &  $H_G(0, k \pi) H$ & 2 \\ 
    $Y^k$               &  $H_G(0,-\pi/2) H_G(0, k \pi) H_G(0,\pi/2) H$ & 4 \\ 
    ${\rm RX}(\theta)$  &  $H_G(0,\theta) H_G(-\theta/2,-\theta/2)$ & 2 \\
    ${\rm RY}(\theta)$  &  $H_G(0,-\pi/2) H_G(0,\theta) H_G(0,\pi/2) H_G(-\theta/2,-\theta/2)$ & 4 \\
    ${\rm RZ}(\theta)$  &  $Z_G(-\theta/2,\theta/2)$  &   1 \\
    \hline
  \end{tabular}
} % scalebox
%\end{table}
~ \\

Users of MuqcsCraft can press a special ``Expand Circuit'' button
to invoke automatic replacement of
the non-core gates in the tables above with core gates,
using generalized gates or not in the resulting expansion.
This expansion makes the circuit deeper but also allows difference highlighting to show
the effects of all gates on the state vector.

\section{Qubit Pairs: The Half-Matrix}\label{sec:HalfMatrix}

\begin{figure}[!thb]
\centering
\includegraphics[width=0.99\columnwidth]{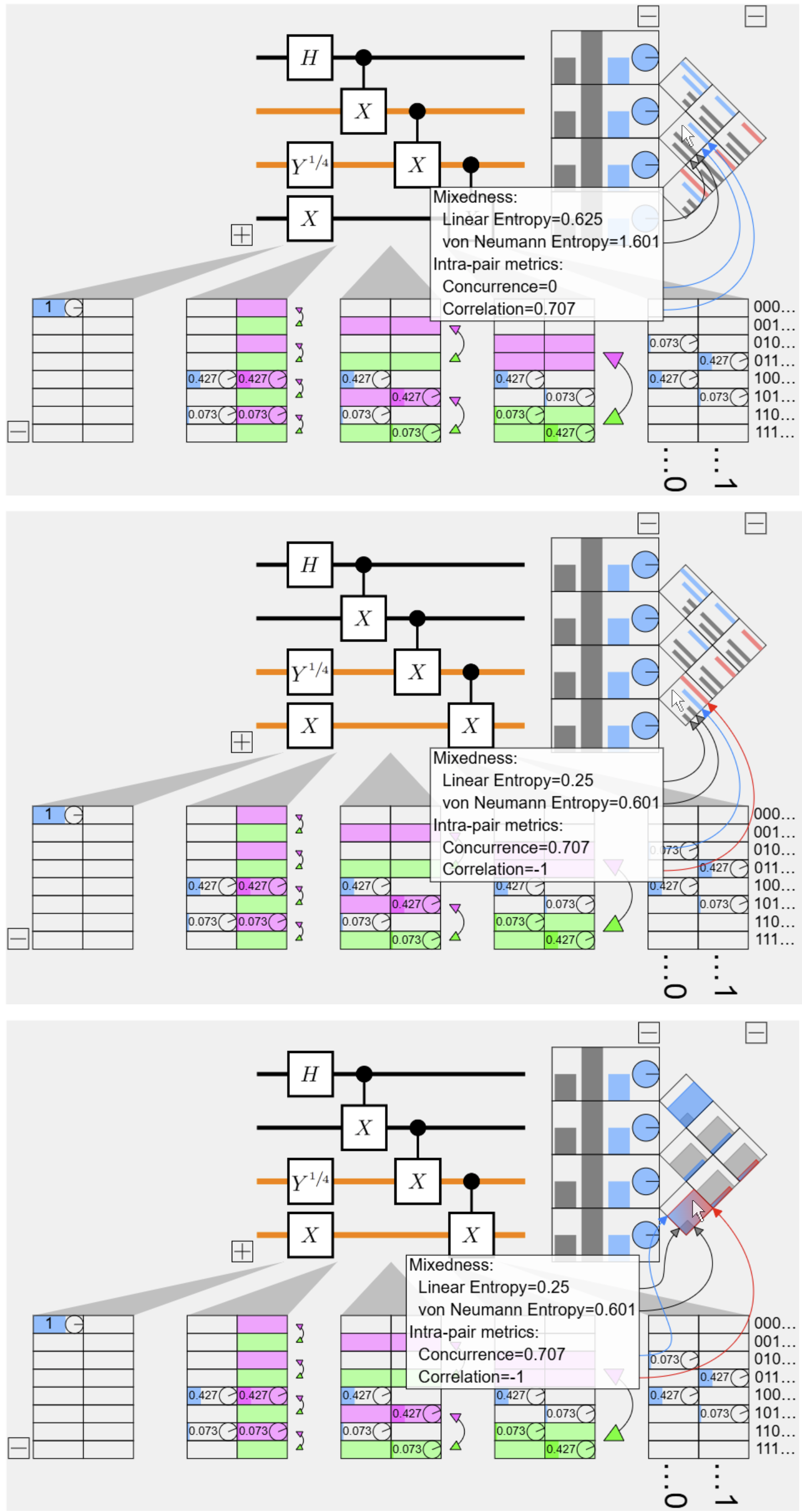}
\caption{
   % A circuit very similar to that in Figure~\ref{fig:ibmAndQuirk}.
   Top: the mouse cursor hovers over the middle-left cell of the half-matrix,
   which corresponds to the two middle qubits, whose wires are highlighted in orange.
   The tooltip reveals that these two qubits have correlation (in the computational basis) of $0.707$, and concurrence $0$.
   Middle: the bottom two qubits have correlation $-1$ and concurrence $0.707$.
   Bottom: the cells of the half-matrix display rectangular glyphs instead of bars,
       which might make it easier to quickly find cells with high or low levels of
       mixedness or correlation / concurrence.
}
\label{fig:halfMatrix}
\end{figure}

A triangular half-matrix is one way to organize pairs of elements,
and has been used in previous visualizations
\cite{wilkinson2005,im2013gplom}
% hartigan1975, elmqvist2008scatterdice, bezerianos2010graphdice, zhao2015matrixwave
not related to quantum computing.
Each cell of our half-matrix corresponds to a pair of qubits, as revealed when the mouse hovers over a cell
and the corresponding qubit wires highlight in orange (Figure~\ref{fig:halfMatrix}).

Figure~\ref{fig:halfMatrix} shows our half-matrix to the right of the circuit diagram. % for visualizing information about pairs of qubits.
The circuit shown is very similar to
that in Figure~\ref{fig:ibmAndQuirk}.
In both figures, the output state
has high probabilities for base states 0111 and 1000,
low probabilities for 0100 or 1011,
and zero for all others.
Our half-matrix in Figure~\ref{fig:halfMatrix}
reveals that the correlation (in the computational basis)
of the top two qubits is $+1$,
that of the bottom two qubits is $-1$,
and that of the middle two qubits is $0.707$.
We might be tempted to conclude that there is a stronger entanglement between the top two and bottom two qubits,
and weaker entanglement between the middle two.
However, correlation is not a proper metric of entanglement, since it depends on the choice of basis.
The half-matrix also shows concurrence \cite{coffman2000distributed} values,
and reveals the top two and bottom two qubits to have a concurrence of 0.707,
while the middle two have a concurrence of zero.
This is not shown at all in the user interfaces of Figure~\ref{fig:ibmAndQuirk}.

In Figure~\ref{fig:halfMatrix}, top and middle, each cell of the half-matrix
shows bars revealing the metrics we chose to display.
Metrics related to mixedness are shown in dark grey,
and those related to the intra-pair relationship are shown in blue if positive or red if negative.
Additional metrics could, of course, be added as desired.
% This could, of course, be easily extended to additional metrics as desired.
We also experimented with showing the metrics as rectangular glyphs \cite{borgo2013}, shown in Figure~\ref{fig:halfMatrix}, bottom.
The two metrics related to mixedness are mapped to the orthogonal dimensions of a dark grey rectangle,
and the two intra-pair metrics are mapped to orthogonal dimensions of another rectangle whose
sides are blue or red to indicate the sign of the metric.
A user accustomed to this visual feedback could plausibly quickly find
rectangles of interest, based on their size and color, in a large half-matrix.

\section{Additional Examples} % Case Studies

% \mjm{Maybe this section should just be called ``additional examples'', since ``case studies'' sounds more serious and we don't have much to say and don't have feedback from users}

\begin{figure}[!thb]
\centering
\includegraphics[width=0.99\columnwidth]{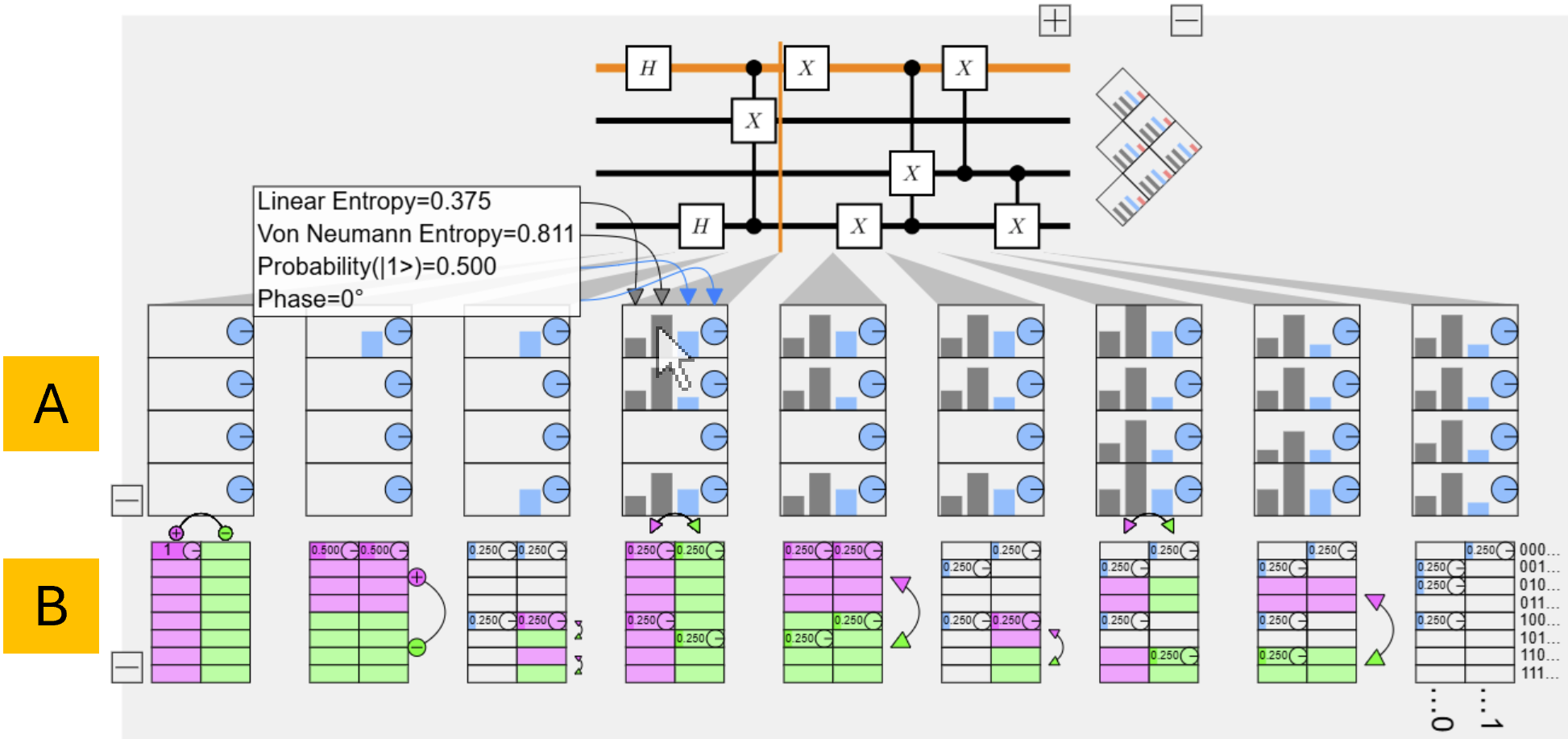}
\caption{A circuit generating a W-4 state.
Below the circuit diagram, for each layer, we see (A) the reduced state for each qubit, and (B) the state vector with difference highlighting.
Which of those two seems more useful for understanding the operation of the circuit?
}
\label{fig:caseStudyW4}
\end{figure}

Figure~\ref{fig:caseStudyW4} shows a circuit that generates a W-4
state, where the four base states with a single 1 in their bitstring
(namely 0001, 0010, 0100, and 1000)
are the only ones with non-zero probability.
For each layer, we see
(Figure~\ref{fig:caseStudyW4}~A) information about the single-qubit reduced states,
and (Figure~\ref{fig:caseStudyW4}~B) \svdh.
Users who learn how to read (B) can use it
to design changes to the circuit,
or to correct small errors in the placement of gates,
since each gate is directly related to the repositioning or rotation of amplitudes.
However, (A) provides much less guidance and establishes a much less direct
relationship between gates and the evolution of the state.
In fact, part of the inspiration for designing the visual feedback in (B)
came from one of us trying to design a circuit in Quirk, through trial-and-error,
to generate a W-4 state, and realizing that gates like a multi-control $X$ gate
(also called generalized Toffoli gate) can move non-zero amplitudes around the state vector,
and that new visual feedback could be used to guide the user in the choice and placement of such gates.

\begin{figure}[!thb]
\centering
\includegraphics[width=0.99\columnwidth]{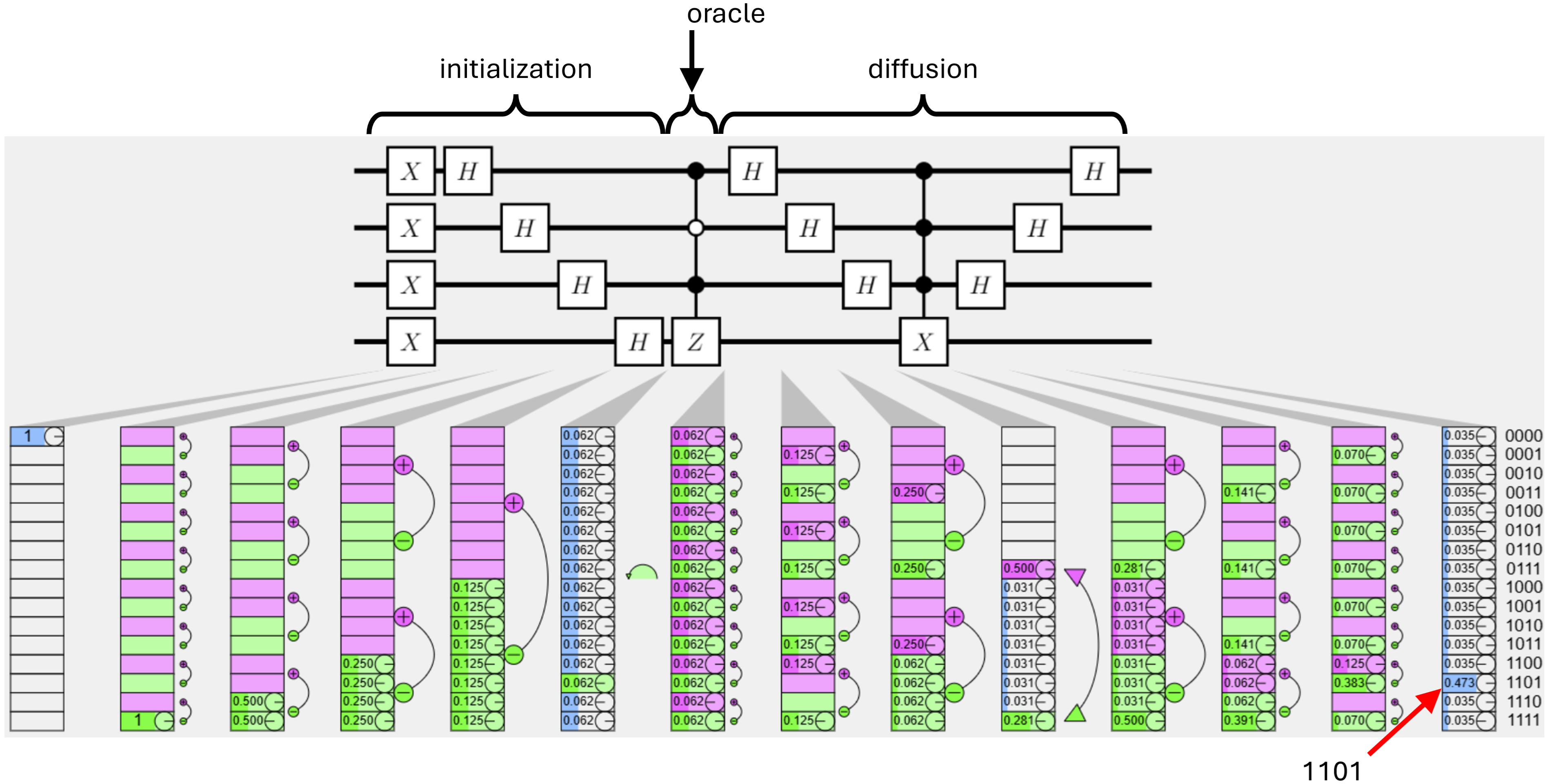}
\caption{A circuit implementing a single iteration of Grover's algorithm, increasing the probability of the amplitude for one base state.
The oracle marks one base state by rotating its amplitude.
}
\label{fig:caseStudyGrover}
\end{figure}

Figure~\ref{fig:caseStudyGrover} shows essential parts of Grover's algorithm:
how the $H$ gates in the initialization spread probability evenly over all base states,
how the oracle rotates a single amplitude,
and how the subsequent $H$ gates concentrate probability onto the marked base state.

\section{Implementation details}

MuqcsCraft \cite{mcguffin2025muqcscraft} is built on top of the open-source Muqcs library \cite{mcguffin2025muqcs}.
The circuit is simulated using a state vector simulator,
where the state vector is updated layer-by-layer using
an efficient implementation of qubit-wise multiplication
(\S3 in \cite{mcguffin2025tutorial}).
To compute quantities related to reduced states,
an efficient partial trace subroutine
(\S6 in \cite{mcguffin2025tutorial})
computes all
single-qubit 2$\times$2 reduced density matrices,
from which qubit probability, phase, purity,
and other quantities are computed.
To populate the half-matrix,
we compute all two-qubit 4$\times$4 reduced density matrices,
from which pairwise concurrences and other quantities are computed.

Like Quirk \cite{gidney2020quirk},
the gates in the circuit are encoded in the URL query string
of the MuqcsCraft web application,
making it possible to bookmark circuits in a web browser
and share hyperlinks to circuits in plain text documents.
Because of this, the web browser's Back button provides a way to undo changes to the circuit.
There are also features to easily export circuits
from MuqcsCraft
to Quirk and to IBM Quantum Composer \cite{ibm2023composer}.

\section{Limitations}

Our \svdh\ requires that each gate appear in its own layer,
is limited to $\approx$ 8 qubits, and only supports a set of ``core'' gates.
Although any unitary gate can be replaced with an equivalent sequence of core gates
whose effects can be visualized, in practice these sequences are sometimes long,
which could impede understanding of their cumulative effect.

% Update: I'm wrong about this; iSWAP and sqrt(SWAP) gates can be decomposed into CNOT and single qubit gates.
%
%Also, although we do visualize single-qubit gates with arbitrary combinations of control and anticontrol qubits, we currently have no way to visualize the effect of other multi-qubit gates like $i$SWAP or $\sqrt{{\rm SWAP}}$.

\section{Related Work}

\subsection{Visualizations for Quantum Computing}

A growing body of work applies principles of information visualization \cite{munzner2014book} and interaction design
to quantum computing, e.g.,
showing higher-level blocks in a circuit \cite{wen2023quantivine},
allowing users to sketch circuits \cite{arawjo2022},
visualizing and manipulating tensor networks \cite{kissinger2015quantomatic},
visualizing noise \cite{ruan2023vacsen}
and graph states \cite{miller2021},
or enabling a better understanding of
Grover's algorithm \cite{norrie2024qgrover},
Shor's algorithm \cite{tao2017shorvis},
quantum machine learning \cite{ruan2024violet},
and variational quantum algorithms (VQAs) \cite{rudolph2021orqviz}.

% We now review some of the more specific work related to visualization of circuits.
We now review topics related more specifically to our current work.

\subsection{Visualization of circuits}

Compared to previous work,
MuqcsCraft is the only graphical circuit simulator
that highlights differences in the state vector from one layer to the next,
as well as the only to display a triangular half-matrix,
and the only one that computes and shows pairwise concurrence values
to aid in understanding entanglement.
Below we discuss previous work and some of their specific features.

\subsubsection{Popular graphical interfaces}

% Some software simulators,  such as Quirk, allow the user to see each qubit's local state visualized in a small 3-dimensional Bloch sphere.
% On the other hand, IBM Quantum Composer shows each qubit's local state using a ``phase disk'' glyph, showing a probability, phase angle $\phi$, and purity.

IBM Quantum Composer \cite{ibm2023composer}
and Quirk \cite{gidney2020quirk} are both web-based, like MuqcsCraft.
Quantum Composer is limited to 4 qubits without an account and is closed source,
whereas Quirk and MuqcsCraft support 16 qubits and are both open source.
Quantum Composer lacks support for anticontrol qubits,
lacks support for control qubits on certain gates ($T$ and $S$),
supports a maximum of one control qubit on most gates,
and requires several clicks to add this control qubit,
whereas Quirk and MuqcsCraft allow arbitrary control and anticontrol qubits to be added to any gate
with a simple drag-and-drop.
Quantum Composer also has no support for continuously varying the parameters of gates,
whereas Quirk supports animated rotation gates,
and MuqcsCraft allows parameters of gates to be smoothly varied by dragging with the mouse
which triggers continuously updated visual feedback.

% Asked on Discord:
% michael mcguffin — 23/03/2024 18:22
% Quelle interface drag-and-drop ou interface graphique utilisez-vous pour définir des circuits?  S.v.p. réagir (thumbs up) aux choix suivants ou ajouter un choix en commentaire:
% IBM Quantum Composer https://quantum.ibm.com/composer/files/new
%     13 votes, including by mcguffin
% Quirk https://algassert.com/quirk
%     1 vote, by mcguffin

% We informally surveyed a quantum computing student club and found that 12 out of 12 members used IBM Quantum Composer to define circuits, and none used Quirk.
% Informal feedback indicates that some users find Quirk's user interface confusing.

To visualize the evolution of the single-qubit reduced states,
both Quantum Composer and Quirk allow widgets to be drag-and-dropped onto the circuit
between each layer of gates,
however MuqcsCraft allows the visibility of reduced states for all layers to be toggled with a single click.

To visualize the evolution of the state vector,
Quantum Composer allows the user to step through the circuit, one layer at a time,
while a single colored barchart of the state vector is updated.
Quirk and MuqcsCraft both display the state vector with wrapping to save space
(and in MuqcsCraft, wrapping is an option that can be toggled).
Quirk allows state vector widgets to be drag-and-dropped onto the circuit
between each layer of gates.
MuqcsCraft instead allows the visibility of state vectors for all layers to be toggled with a single click,
and is the only package to highlight the differences from one layer to the next.

Comparing the two open-source packages,
Quirk is over 40k lines of code versus MuqcsCraft's less than 10k
making it easier to study and understand.
Quirk uses GPU acceleration to simulate circuits faster,
while MuqcsCraft avoids the complexity of GPU programming
and nevertheless simulates circuits of 16 qubits or less in under 10 ms per gate on a 2022 laptop.

\subsubsection{Academic work on circuit visualization}

Rainbow boxes \cite{lamy2019rainbow}
represent qubits with rectangles,
where height shows probability, color shows phase,
and entanglement is indicated by merging rectangles,
however this can only be done for adjacent qubits,
and makes no distinction between qubits that are
partially versus maximally entangled.

QuFlow \cite{lin2018quflow} shows causal relationships between amplitudes from one layer to the next,
and
QuantumEyes \cite{ruan2023quantumeyes} has a Probability Summary View that shows the evolution of
probabilities of base states layer-by-layer.
However, neither of these use
wrapping to make the state vectors consume less
vertical space,
making them less scalable than
Quirk and MuqcsCraft.
% QuantumEyes offers demo but only 2 fixed circuits.  Update: their demo doesn't even fully work anymore.
In addition, there is no obvious way to modify
these techniques to support wrapping.

% don't bother discussing QCVis (Williams 2021) ?

Also, unlike these previous works,
we provide the source code for MuqcsCraft
and a working interactive online demo.

%\begin{figure}[!thb]
%\centering
%\includegraphics[width=0.99\columnwidth]{fig/table3.png}
%\caption{TODO This figure needs to be updated and maybe even removed.
%}
%\label{fig:table}
%\end{figure}

\subsection{Extensions to the Bloch sphere}

Methods for embedding sets of quantum states in space
can be developed through
the perspective of abstract algebra and topology,
discussed in depth by Bengtsson and {\.Z}yczkowski \cite{bengtsson2017geometry}.
Previous works have extended the Bloch sphere
to two qubits \cite{gamel2016,wie2020,chang2024},
e.g., to better understand entanglement,
and even to multiple qubits
\cite{altepeter2009,makela2010,bley2024}, % aharonov2025
however these extensions require 3D geometry
that is difficult to visualize,
and/or have limited scalability to many qubits.
% \mjm{Je n'arrive pas à comprendre \cite{aharonov2025} qui est seulement sur arxiv et n'a pas
% encore passé par un processus de ``peer review'', donc je me demande si on devrait vraiment le citer.
% Explication d'un IA de \cite{aharonov2025}: \url{https://claude.ai/share/0ba2f25b-1f1b-4e8e-9d98-7522e6806b07}}

In contrast, the triangular half-matrix used in MuqcsCraft
is purely 2D, and occupies an area that scales quadratically with
the number of qubits, allocating one cell for each pair of qubits
to show information about how the pair are related.

% discuss this too \cite{ruan2023venus} ?

\section{Conclusions and Future Directions}

We presented two novel visualization techniques for quantum circuits,
\svdh\ and the triangular half-matrix,
demonstrated in our open-source MuqcsCraft software.
Difference highlighting uses a small set of visual primitives
to illustrate the effect of each gate,
and is visually universal over single-qubit gates with control and anticontrol qubits,
if we substitute certain gates with equivalent sequences.
The half-matrix provides pairwise information about qubits,
revealing entanglement and correlation patterns that are
not visible in common single-qubit reduced state displays.
This work is a further step toward tools for understanding and designing quantum algorithms.

In future work,
difference highlighting might be modified to
be applicable to subsets of qubits, making our system more scalable to large circuits;
% or extended to work with multi-qubit gates like $i$SWAP or $\sqrt{{\rm SWAP}}$;
or extended to illustrate the aggregated effect of multi-qubit gates or higher-level blocks of gates.
For example, visualizing the aggregated effect of the diffuser in Grover's algorithm,
iterated a few times, would help to illustrate the algorithm.
It could also be fruitful to consider changes of basis in how the state vector is visualized,
making certain gates easier to understand.
The half-matrix might be modified to allow wires
to be reordered \cite{behrisch2016}
to cluster interesting subsets of qubits,
or only show the most interesting pairs of qubits and occupy less space.

% state vector: ability to change basis, to make certain gates easier to understand

% half-matrix: ability to reorder wires to cluster interesting subsets of qubits

% hardware acceleration ?

% \mjm{Generated by AI:}
% While our techniques are currently limited to circuits with approximately 8 qubits due to screen space constraints, hybrid approaches could selectively apply difference highlighting to subsets of qubits in larger circuits, or use hierarchical visualization to show both local and global circuit behavior.
% The visual vocabulary could be extended to directly support multi-qubit gates like quantum Fourier transform components, potentially through new visual primitives or animated transitions.

% if have a single appendix:
%\appendix[Proof of the Zonklar Equations]
% or
%\appendix  % for no appendix heading
% do not use \section anymore after \appendix, only \section*
% is possibly needed

% use appendices with more than one appendix
% then use \section to start each appendix
% you must declare a \section before using any
% \subsection or using \label (\appendices by itself
% starts a section numbered zero.)
%

%\appendices
%\section{Proof of the First Zonklar Equation}
%Appendix one text goes here.

% you can choose not to have a title for an appendix
% if you want by leaving the argument blank
%\section{}
%Appendix two text goes here.

\ifthenelse{\boolean{IncludeAppendix}}{

\appendix[Calibration of HoloLens]

% \subsection{Calibration of HoloLens}

% \subsubsection{Calibration}

We implemented calibration with a homography
matrix that transforms ...

}{} % ifthenelse

% use section* for acknowledgment
\ifCLASSOPTIONcompsoc
  % The Computer Society usually uses the plural form
  \section*{Acknowledgments}
\else
  % regular IEEE prefers the singular form
  \section*{Acknowledgment}
\fi

Thanks
to Thomas R. Bromley who gave advice and implemented a version of the pairwise concurrence computation for the half-matrix,
and to Olivier Landon-Cardinal who gave suggestions for how to evaluate the software prototype.
This research was supported financially by NSERC and by Alicia Servera.

% Can use something like this to put references on a page
% by themselves when using endfloat and the captionsoff option.
\ifCLASSOPTIONcaptionsoff
  \newpage
\fi

% trigger a \newpage just before the given reference
% number - used to balance the columns on the last page
% adjust value as needed - may need to be readjusted if
% the document is modified later
%\IEEEtriggeratref{8}
% The "triggered" command can be changed if desired:
%\IEEEtriggercmd{\enlargethispage{-5in}}

% references section

% The IEEEtran BibTeX style support page is at:
% http://www.michaelshell.org/tex/ieeetran/bibtex/
%\bibliographystyle{IEEEtran}
% argument is your BibTeX string definitions and bibliography database(s)
%\bibliography{IEEEabrv,../bib/paper}

\bibliography{main}

% Generated by IEEEtran.bst, version: 1.14 (2015/08/26)
\begin{thebibliography}{10}
\providecommand{\url}[1]{#1}
\csname url@samestyle\endcsname
\providecommand{\newblock}{\relax}
\providecommand{\bibinfo}[2]{#2}
\providecommand{\BIBentrySTDinterwordspacing}{\spaceskip=0pt\relax}
\providecommand{\BIBentryALTinterwordstretchfactor}{4}
\providecommand{\BIBentryALTinterwordspacing}{\spaceskip=\fontdimen2\font plus
\BIBentryALTinterwordstretchfactor\fontdimen3\font minus \fontdimen4\font\relax}
\providecommand{\BIBforeignlanguage}[2]{{%
\expandafter\ifx\csname l@#1\endcsname\relax
\typeout{** WARNING: IEEEtran.bst: No hyphenation pattern has been}%
\typeout{** loaded for the language `#1'. Using the pattern for}%
\typeout{** the default language instead.}%
\else
\language=\csname l@#1\endcsname
\fi
#2}}
\providecommand{\BIBdecl}{\relax}
\BIBdecl

\bibitem{mcguffin2025tutorial}
M.~J. McGuffin, J.-M. Robert, and K.~Ikeda, ``How to write a simulator for quantum circuits from scratch: A tutorial,'' 2025, https://arxiv.org/abs/2506.08142.

\bibitem{ibm2023composer}
IBM, ``{IBM} quantum composer,'' 2023, https://quantum.cloud.ibm.com/composer.

\bibitem{gidney2020quirk}
C.~Gidney, ``Quirk,'' 2020, https://algassert.com/quirk.

\bibitem{coffman2000distributed}
V.~Coffman, J.~Kundu, and W.~K. Wootters, ``Distributed entanglement,'' \emph{Physical Review A}, vol.~61, no.~5, p. 052306, 2000.

\bibitem{mcguffin2025muqcscraft}
M.~J. McGuffin, ``{MuqcsCraft}: A web-based graphical simulator and visualizer for quantum circuits,'' 2025, https://github.com/MJMcGuffin/MuqcsCraft.

\bibitem{munzner2014book}
T.~Munzner, \emph{Visualization Analysis and Design}.\hskip 1em plus 0.5em minus 0.4em\relax CRC press, 2014.

\bibitem{tan2023}
S.~Tan, C.~Lai, X.~L. Zhang, and X.~Yuan, ``An empirical guide for visualization consistency in multiple coordinated views,'' in \emph{IEEE Pacific Visualization Symposium (PacificVis)}, 2023, pp. 31--40.

\bibitem{mottonen2004}
M.~M{\"o}tt{\"o}nen, J.~J. Vartiainen, V.~Bergholm, and M.~M. Salomaa, ``Quantum circuits for general multiqubit gates,'' \emph{Physical review letters}, vol.~93, no.~13, p. 130502, 2004.

\bibitem{shende2006}
V.~Shende, S.~Bullock, and I.~Markov, ``Synthesis of quantum-logic circuits,'' \emph{IEEE Transactions on Computer-Aided Design of Integrated Circuits and Systems}, vol.~25, no.~6, pp. 1000--1010, 2006.

\bibitem{wilkinson2005}
L.~Wilkinson, A.~Anand, and R.~Grossman, ``Graph-theoretic scagnostics,'' in \emph{IEEE Symp. Information Visualization (InfoVis)}, 2005.

\bibitem{im2013gplom}
J.-F. Im, M.~J. McGuffin, and R.~Leung, ``{GPLOM}: the generalized plot matrix for visualizing multidimensional multivariate data,'' \emph{IEEE Transactions on Visualization and Computer Graphics (TVCG)}, 2013.

\bibitem{borgo2013}
R.~Borgo, J.~Kehrer, D.~H. Chung, E.~Maguire, R.~S. Laramee, H.~Hauser, M.~Ward, and M.~Chen, ``Glyph-based visualization: Foundations, design guidelines, techniques and applications,'' in \emph{Eurographics STARs (State of the Art Reports)}, 2013, pp. 39--63.

\bibitem{mcguffin2025muqcs}
M.~J. McGuffin, ``Muqcs.js: {McGuffin's} useless quantum circuit simulator,'' 2025, https://github.com/MJMcGuffin/muqcs.js.

\bibitem{wen2023quantivine}
Z.~Wen, Y.~Liu, S.~Tan, J.~Chen, M.~Zhu, D.~Han, J.~Yin, M.~Xu, and W.~Chen, ``Quantivine: A visualization approach for large-scale quantum circuit representation and analysis,'' \emph{IEEE Transactions on Visualization and Computer Graphics (TVCG)}, 2023.

\bibitem{arawjo2022}
I.~Arawjo, A.~DeArmas, M.~Roberts, S.~Basu, and T.~Parikh, ``Notational programming for notebook environments: A case study with quantum circuits,'' in \emph{ACM Symp. User Interface Software and Technology (UIST)}, 2022, pp. 1--20.

\bibitem{kissinger2015quantomatic}
A.~Kissinger and V.~Zamdzhiev, ``Quantomatic: A proof assistant for diagrammatic reasoning,'' in \emph{CADE-25: 25th International Conference on Automated Deduction}.\hskip 1em plus 0.5em minus 0.4em\relax Springer, 2015, pp. 326--336.

\bibitem{ruan2023vacsen}
S.~Ruan, Y.~Wang, W.~Jiang, Y.~Mao, and Q.~Guan, ``{VACSEN}: A visualization approach for noise awareness in quantum computing,'' \emph{IEEE Transactions on Visualization and Computer Graphics (TVCG)}, vol.~29, no.~1, pp. 462--472, 2023.

\bibitem{miller2021}
M.~Miller and D.~Miller, ``{GraphStateVis}: Interactive visual analysis of qubit graph states and their stabilizer groups,'' in \emph{IEEE Int. Conf. Quantum Computing and Engineering (QCE)}, 2021, pp. 378--384, https://github.com/graphstatevis/app.

\bibitem{norrie2024qgrover}
S.~Norrie, A.~Estey, H.~M{\"u}ller, and U.~Stege, ``{QGrover}: Teaching grover's algorithm through visual exploration,'' in \emph{IEEE Int. Conf. Quantum Computing and Engineering (QCE)}, vol.~3, 2024, pp. 17--24.

\bibitem{tao2017shorvis}
Z.~Tao, Y.~Pan, A.~Chen, and L.~Wang, ``{ShorVis}: A comprehensive case study of quantum computing visualization,'' in \emph{Int. Conf. Virtual Reality and Visualization (ICVRV)}.\hskip 1em plus 0.5em minus 0.4em\relax IEEE, 2017, pp. 360--365.

\bibitem{ruan2024violet}
S.~Ruan, Z.~Liang, Q.~Guan, P.~Griffin, X.~Wen, Y.~Lin, and Y.~Wang, ``{VIOLET}: Visual analytics for explainable quantum neural networks,'' \emph{IEEE Transactions on Visualization and Computer Graphics (TVCG)}, vol.~30, no.~6, pp. 2862--2874, 2024.

\bibitem{rudolph2021orqviz}
M.~S. Rudolph, S.~Sim, A.~Raza, M.~Stechly, J.~R. McClean, E.~R. Anschuetz, L.~Serrano, and A.~Perdomo-Ortiz, ``{ORQVIZ}: Visualizing high-dimensional landscapes in variational quantum algorithms,'' 2021, https://arxiv.org/abs/2111.04695.

\bibitem{lamy2019rainbow}
J.-B. Lamy, ``Dynamic software visualization of quantum algorithms with rainbow boxes,'' in \emph{International Conference on Information Visualization Theory and Applications (IVAPP)}, 2019.

\bibitem{lin2018quflow}
S.~Lin, J.~Hao, and L.~Sun, ``{QuFlow}: Visualizing parameter flow in quantum circuits for understanding quantum computation,'' in \emph{IEEE Scientific Visualization Conference (SciVis)}, 2018, pp. 37--41.

\bibitem{ruan2023quantumeyes}
S.~Ruan, Q.~Guan, P.~Griffin, Y.~Mao, and Y.~Wang, ``{QuantumEyes}: Towards better interpretability of quantum circuits,'' \emph{IEEE Transactions on Visualization and Computer Graphics (TVCG)}, 2023, https://quantumeyes.github.io/.

\bibitem{bengtsson2017geometry}
I.~Bengtsson and K.~{\.Z}yczkowski, \emph{Geometry of quantum states: an introduction to quantum entanglement}, 2nd~ed.\hskip 1em plus 0.5em minus 0.4em\relax Cambridge University Press, 2017.

\bibitem{gamel2016}
O.~Gamel, ``Entangled bloch spheres: Bloch matrix and two-qubit state space,'' \emph{Physical Review A}, vol.~93, no.~6, p. 062320, 2016.

\bibitem{wie2020}
C.-R. Wie, ``Two-qubit bloch sphere,'' \emph{Physics}, vol.~2, no.~3, pp. 383--396, 2020.

\bibitem{chang2024}
L.-H.~H. Chang, S.~Roccaforte, Z.~Xu, and P.~Cadden-Zimansky, ``Geometric visualizations of single and entangled qubits,'' \emph{American Journal of Physics}, vol.~92, no.~7, pp. 528--537, 2024.

\bibitem{altepeter2009}
J.~B. Altepeter, E.~R. Jeffrey, M.~Medic, and P.~Kumar, ``Multiple-qubit quantum state visualization,'' in \emph{Conf. Lasers and Electro-Optics and Int. Quantum Electronics Conf. (CLEO/IQEC)}.\hskip 1em plus 0.5em minus 0.4em\relax IEEE, 2009, pp. 1--2.

\bibitem{makela2010}
H.~M{\"a}kel{\"a} and A.~Messina, ``N-qubit states as points on the bloch sphere,'' \emph{Physica Scripta}, vol. 2010, no. T140, p. 014054, 2010.

\bibitem{bley2024}
J.~Bley, E.~Rexigel, A.~Arias, N.~Longen, L.~Krupp, M.~Kiefer-Emmanouilidis, P.~Lukowicz, A.~Donhauser, S.~K{\"u}chemann, J.~Kuhn, and A.~Widera, ``Visualizing entanglement in multiqubit systems,'' \emph{Physical Review Research}, vol.~6, no.~2, p. 023077, 2024.

\bibitem{behrisch2016}
M.~Behrisch, B.~Bach, N.~H. Riche, T.~Schreck, and J.-D. Fekete, ``Matrix reordering methods for table and network visualization,'' \emph{Computer Graphics Forum}, 2016, http://matrixreordering.dbvis.de/.

\end{thebibliography}
\bibliographystyle{IEEEtran}

% biography section
% 
% If you have an EPS/PDF photo (graphicx package needed) extra braces are
% needed around the contents of the optional argument to biography to prevent
% the LaTeX parser from getting confused when it sees the complicated
% \includegraphics command within an optional argument. (You could create
% your own custom macro containing the \includegraphics command to make things
% simpler here.)
%\begin{IEEEbiography}[{\includegraphics[width=1in,height=1.25in,clip,keepaspectratio]{mshell}}]{Michael McGuffin}
% or if you just want to reserve a space for a photo:

%\begin{IEEEbiography}{Michael J. McGuffin}
%Biography text here.
%\end{IEEEbiography}

\begin{IEEEbiography}[{\includegraphics[width=1in]{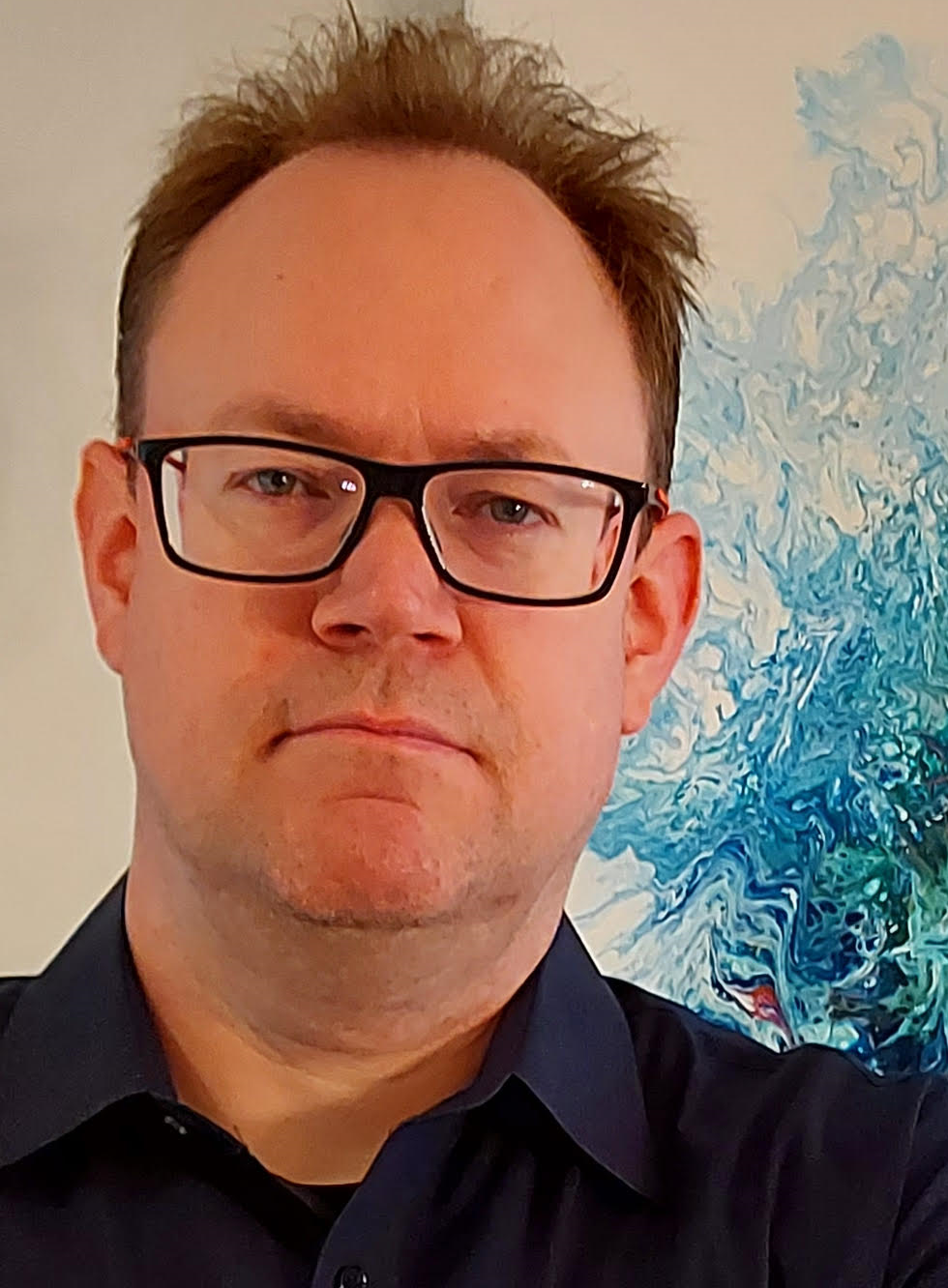}}]{Michael J. McGuffin} is a full professor at ETS, a French-language engineering school in Montreal, Canada, where his students do research in visualization and HCI.
In 2009, his paper at the IEEE Information Visualization Conference (InfoVis 2009) received an Honorable Mention.
\end{IEEEbiography}

\begin{IEEEbiography}[{\includegraphics[width=1in]{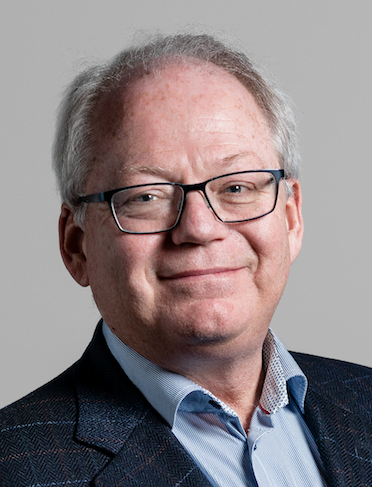}}]{Jean-Marc Robert} is a full professor at ETS
where his students do research in IT security and quantum cryptography. His recent interests include quantum computing and quantum information theory. In 1985, he completed his Master's degree, formalizing the concept of privacy amplification, which led to the second paper ever published on quantum cryptography.
\end{IEEEbiography}

% if you will not have a photo at all:
%\begin{IEEEbiographynophoto}{John Doe}
%Biography text here.
%\end{IEEEbiographynophoto}

% insert where needed to balance the two columns on the last page with
% biographies
%\newpage

%\begin{IEEEbiographynophoto}{Jane Doe}
%Biography text here.
%\end{IEEEbiographynophoto}

% You can push biographies down or up by placing
% a \vfill before or after them. The appropriate
% use of \vfill depends on what kind of text is
% on the last page and whether or not the columns
% are being equalized.

%\vfill

% Can be used to pull up biographies so that the bottom of the last one
% is flush with the other column.
%\enlargethispage{-5in}

\end{document}